\documentclass[12pt,onecolumn,draftclsnofoot]{IEEEtran}
\IEEEoverridecommandlockouts
\usepackage{color}

\usepackage{times}
\usepackage[final]{graphicx}
\usepackage{amsmath,amsfonts}
\usepackage{amsthm}
\usepackage{amssymb,amsbsy}

\usepackage{cite}
\usepackage{multirow}

\usepackage{algorithm}
\usepackage{algpseudocode}

\newcommand{\ignore}[1]{}

\newcommand{\hspp}{\hspace{0.05in} }
\newcommand{\hsppp}{\hspace{0.02in} }

\newsavebox{\savepar}

\newcommand{\ParState}[1]{\parbox[t]{\dimexpr\linewidth-\algorithmicindent}{%
      \setlength{\hangindent}{\algorithmicindent}
      {#1}\strut}}

\begin{document}
\title{Statistical Blockage Modeling and Robustness of Beamforming
in Millimeter Wave Systems}
\author{\normalsize Vasanthan Raghavan, Lida Akhoondzadeh-Asl,
Vladimir Podshivalov, Joakim Hulten, \\
M.\ Ali Tassoudji, Ozge Hizir Koymen, Ashwin Sampath, and Junyi Li \\
Qualcomm Corporate R\&D, USA, \\
Contact E-mail: {\tt vasanthan\_raghavan@ieee.org}
\thanks{A short version of this paper~\cite{vasanth_blockage_spawc2018} 
has been submitted for publication to the
IEEE International Workshop on Signal Processing Advances in Wireless
Communications (SPAWC), Kalamata, Greece, 2018.}
}

\maketitle
\vspace{-10mm}

\begin{abstract}
\noindent
There has been a growing interest in the commercialization of millimeter wave
(mmW) technology as a part of the Fifth-Generation New Radio (5G-NR) wireless
standardization efforts. In this direction, many sets of independent measurement
campaigns show that wireless propagation at mmW carrier frequencies is only
marginally worse than propagation at sub-$6$ GHz carrier frequencies for
small-cell coverage --- one of the most important use-cases for 5G-NR. On the
other hand, the biggest determinants of viability of mmW systems in practice
are penetration and blockage of mmW signals through different materials in the
scattering environment. With this background, the focus of this paper is on
understanding the impact of blockage of mmW signals and reduced spatial
coverage due to penetration through the human hand, body, vehicles, etc.
Leveraging measurements with a $28$ GHz mmW experimental prototype and
electromagnetic simulation studies, we first propose statistical blockage
models to capture the impact of the hand, human body and vehicles. We then
study the time-scales at which mmW signals are disrupted by blockage (hand and
human body). Our results show that these events can be attributed to physical
movements and the time-scales corresponding to blockage are hence on the order
of a few $100$ ms or more. Building on this fundamental understanding, we finally
consider the broader question of robustness of mmW beamforming to handle blockage.
Network densification, subarray switching in a user equipment (UE) designed with
multiple subarrays, fall back mechanisms such as codebook enhancements and
switching to legacy carriers in non-standalone deployments, etc.\ can address
blockage before it leads to a deleterious impact on the mmW link margin.
\end{abstract}



\section{Introduction}
\label{sec1}
The Fifth-Generation New Radio (5G-NR) wireless standardization efforts
are an important component in the successful commercialization of millimeter
wave (mmW) technology~\cite{khan,qualcomm,rappaport,boccardi1} for enhanced
mobile broadband applications. 
In this direction, a number of (multi-institutional) efforts have focussed on the
scope and scale of unfavorableness in wireless propagation at mmW carrier
frequencies relative to sub-$6$ GHz
systems~\cite{3gpp_CM_rel14_38901,5G_whitepaper,802d11_maltsev,metis2020_tcom,vasanth_tap2018}.
These studies show that while propagation losses at mmW frequencies are
typically higher than with sub-$6$ GHz systems in both indoor and outdoor
settings, these losses are not\footnote{In particular, for non-line-of-sight
(NLOS) links,~\cite{vasanth_tap2018} shows a nominal path loss degradation at
$29$ GHz relative to $2.9$ GHz of $5.3$ dB, $2.76$ dB and $3.00$ dB at a
coverage distance of $d = 25$ m in an indoor office setting, $d = 200$ m in
an Urban Micro setting, and $d = 100$ m in a shopping mall setting, respectively.
This degradation is computed as $\left( {\sf PLE}\big|_{29 \hsppp {\sf GHz}} -
{\sf PLE}\big|_{2.9 \hsppp {\sf GHz}} \right) \cdot 10 \log_{10}(d)$ dB at
a coverage distance of $d$ m where ${\sf PLE}\big|_{29 \hsppp {\sf GHz}}$ and
${\sf PLE}\big|_{2.9 \hsppp {\sf GHz}}$ denote the path loss exponent at the
two carrier frequencies in the scenario of interest.} significantly worse at
the mmW regime. These additional propagation losses can be overcome by array
gains reaped from the use of larger antenna arrays (at both
ends)~\cite{rusek,hur,sun,brady_tcom,oelayach,raghavan_jstsp,vasanth_jsac2017,balanis},
increased effective isotropic radiated power (EIRP) levels~\cite{fcc_ruling},
and system design that aids in opportunistic signaling by leveraging time,
frequency and space diversity~\cite{roh,rangan,ghosh,vasanth_comm_mag_16}.

Nevertheless, as
the 5G-NR design process marches towards the accelerated schedule of early
commercial deployments, important aspects that determine the viability of
mmW technology in practice, such as outdoor-to-indoor penetration and
blockage, have to be addressed carefully.
Outdoor-to-indoor penetration through different types of residential/office
materials has been studied extensively; see
e.g.,~\cite[Appendix E]{metis2020_tcom},~\cite{vasanth_tap2018}. These works
report that the reflection response and penetration loss are a function of
the material property, frequency, polarization and incident angle, and
significantly deep signal reception notches spread over several GHz of the
spectrum are observed. Such an observation motivates the need for system
designs that support {\em both} frequency and spatial diversity.

In the context of spatial diversity, given the use of large antenna arrays and
the diminishing beamwidths of the directional beams with antenna
dimensions~\cite{oelayach,raghavan_jstsp,vasanth_jsac2017,balanis}, mmW systems are
susceptible to signal blockage much more than sub-$6$ GHz systems are. In
particular, mmW systems are susceptible to {\em self-blockage}, which is shadowing
from the user itself in the form of hand blocking and blockage from other body
parts. This can cause a complete blockage of the user equipment (UE) antennas
depending on the antenna position relative to the hand. In addition, there are
blockages from the environment around the UE in the form of buildings, foliage,
or other obstructions ({\em static blockage}) and humans, vehicles, or moving
obstructions ({\em dynamic blockage}).

\noindent {\bf \em \underline{Prior Work:}} In prior work, a human blockage model
has been included in the $802.11$(ad) $60$
GHz wireless standardization efforts~\cite[Sections 3.3.8, 3.5.7, 5.3.9, 8]{802d11_maltsev}.
This model
captures the probability of a cluster blockage event and the distribution function
of power attenuation for these events, both via ray-tracing studies. Wideband $60$
GHz human blockage measurements over a
$3$ GHz bandwidth has been performed in~\cite{peter_et_al} and the authors study
the comparative model fits between the double knife edge diffraction (DKED) and the
uniform theory of diffraction (UTD) modeling frameworks showing that the DKED
framework underestimates blockage loss and the UTD framework overestimates it.

The Mobile and wireless communications Enablers for the Twenty-twenty Information
Society (METIS) project has proposed a human blockage model based on the DKED
framework in~\cite[pp.\ 39-41, 160-162]{metis2020_tcom}. A blockage model is proposed
by the 3GPP Rel.\ 14 channel modeling document~\cite[pp.\ 53-57]{3gpp_CM_rel14_38901}
for mmW system modeling under two variants: a stochastic variant (Option A)
and a map-based variant (Option B). Both these variants assume a $30$ dB flat loss for
self-blockage. A modified version of the METIS model based on $73$ GHz human blockage
measurements using horn antennas has been proposed to account for directional transmissions
in~\cite{maccartney_vtc2016,maccartney_2017,5G_whitepaper} and these studies show
that human blockage could cause signal attenuation on the order of $30$-$40$ dB
depending on the distance between the human and the transmitter/receiver. Another
work that studies UE antenna modeling in form-factor UE designs at $15$ GHz from
a blockage consideration is~\cite{zhao_ericsson}. This study illustrates performance
losses with the hand phantom model and recommends the use of subarray diversity to
overcome these losses. The readers are pointed to~\cite{tap_overview1,tap_overview2}
for recent studies on design tradeoffs of 5G antenna arrays with form-factor
considerations.

\noindent {\bf \em \underline{Contributions:}} With this backdrop, we first note
that most of the prior works focus specifically on human blockers in a low-mobility
indoor setting with short transmit-receive distances using horn antennas for
measurements. On the other hand, the use of a phased array ($2$-$8$ antennas)
at the UE end implies that the beamwidth\footnote{For example, a beam with
progressive phase shifts (or a constant phase offset) over a $4$ element linear
array is expected to lead to a $3$ dB-beamwidth of $\approx 25^{\sf o}$, whereas
a horn antenna typically has a $3$ dB-beamwidth on the order of
$7.5^{\sf o}$-$15^{\sf o}$ at mmW frequencies.} at the UE side is expected to be
much larger than that seen with a horn antenna. Such differences can lead to a
significant variation between the blockage modeling experiments with a form-factor
UE design and horn antenna measurements. In general, while form-factor UE design-based
blockage modeling studies provide the {\em gold standard} in terms of understanding
the implications of blockage at the UE end, such studies are currently difficult to
obtain due to the still ongoing design, manufacture, development and testing of mmW
technology supporting UEs/chipset solutions.

We propose to address these shortfalls by an intelligent mix of measurements with
a form-factor UE prototype (as reported in~\cite{vasanth_comm_mag_16}) and
electromagnetic simulation studies 
for form-factor designs where such studies can supplant measurement-based insights.
In particular, simulation studies are useful to understand the loss in spatial coverage
with blockage. On the other hand, mmW measurements provide the best estimate
for loss in signal strength as well as time-scales at which blockage disruptions
happen. In this work, we provide a complementary mix of simulation studies and
measurement studies for both self- and dynamic blockage.

In this direction, we first study self-blockage by considering electromagnetic
simulations of antennas at $28$ and $60$ GHz in the proximity of the hand and
identifying the spatial regions corresponding to signal blockage with different
user grips. These simulation studies illustrate the blockage of a large spatial
region in the UE's local coordinate system depending on whether the UE is held
in a {\em Portrait} or {\em Landscape} mode. We then study the loss incurred by
the hand with different grip experiments using the $28$ GHz prototype reported
in~\cite{vasanth_comm_mag_16}. In contrast to prior work that shows high
self-blockage losses ($30$ to $40$ dB), our studies show that a median loss of
$15$ dB is incurred by the hand even in the most pessimistic scenario of a hard
hand grip. The beamwidth differences between horn antennas and a phased array
design is likely to account for such wide discrepancies.

To model dynamic blockage, we conduct simulation studies to capture the impact of
objects at the UE end in the form of angular regions blocked and losses incurred
with a DKED model. We study the efficacy of the simulated loss data with measurement
studies sing the $28$ GHz prototype. Our studies show that though there are some
discrepancies between simulated loss and true measurements, simulated data can
offer a reasonable first-order estimate of blockage losses and can thus be
useful for scenarios like vehicular applications, where loss estimation with
measurements is considerably more complicated and difficult. These studies lead
to the proposal of a statistical blockage model that has many attractive
properties: i) parsimonious (captured by a small number of model parameters),
ii) efficacious (captures the real impact of blockages), and iii) computationally
efficient (easily useable in a system simulator framework) in studying the
performance of mmW systems.

We then consider the question of time-scales at which mmW signals are disrupted
due to blockage. Such time-scale estimation is important to understand the scope of
mmW beamforming solutions, their robustness/stability and the nature of mitigation
mechanisms to handle blockage without serious link degradation. With some prototype
studies, we show that these time-scales can be attributed to physical movements
of the source of blockage (hand in the case of self-blockage, humans/vehicles
in the case of dynamic blockage, etc.). Thus, the dynamics of these blockers capture
the time-scales at which mmW signals get disrupted and measurement studies show that
they are on the order of a few $100$ ms (or more). Given the sub-ms (or a few ms)
effective latencies targeted by 5G-NR for beam/subarray switching, it appears that
the deleterious impact of blockages can be addressed by a robust beam management
procedure at the PHY layer level. In terms of PHY layer solutions, network
densification, design of multiple subarrays and capability to switch beams/subarrays
at the UE end, and alternate fall back mechanisms could address these challenges.

\noindent {\bf \em \underline{Organization:}} This paper is organized as
follows. Sections~\ref{sec2} and~\ref{sec3} consider self- and dynamic blockage
from both electromagnetic simulation studies and measurement perspectives,
and simple statistical models are proposed to capture the effect of these
blockages. Section~\ref{sec4} considers the question of time-scales at which
blockage events happen. 
Section~\ref{sec5} proposes multiple approaches to combat the effect of
blockages and Section~\ref{sec6} concludes the paper.

\section{Self-Blockage}
\label{sec2}
The focus of this section is on understanding the impact of self-blockage in
terms of the spatial/angular coverage lost as well as the loss incurred over the
blocked angles.

\subsection{Loss in Spatial/Angular Coverage}
\label{sec2a}

Objects that are electrically small at microwave frequencies become electrically
large at mmW frequencies, and small objects (which have the size of a few mm's)
located in the proximity of the antennas affect the antenna performance and
deteriorate both their efficiencies and radiation patterns. For example, antennas
placed on the display side can be affected by the liquid crystal display (LCD)
shielding, LCD glass, component shields, as well as other objects such as camera(s),
speaker, microphone, sensors, etc.

To investigate the effect of the antennas' surroundings on its performance (especially
its radiation pattern), the antenna module is placed over a simplified model of a
UE (corresponding to a typical size of $60 \times 130$ mm$^2$) and studied in an
electromagnetic simulation framework. The model of the UE simulated consists of
several layers of materials: Glass with a thickness of $1$ mm, LCD shielding which lies beneath
the glass and extends $15$ mm from the edge of the glass, and the FR-4 board\footnote{Note
that ``FR-4 (or FR4) is a grade designation assigned to glass-reinforced epoxy laminate
sheets, tubes, rods and printed circuit boards. FR-4 is a composite material composed
of woven fiberglass cloth with an epoxy resin binder that is flame resistant.''
See {\tt https://en.wikipedia.org/wiki/FR-4} for more details.} with a thickness of
$0.8$ mm that is separated by an $8$ mm air gap from the LCD shielding. Also, a
battery and few shielding boxes of random sizes are placed over the printed circuit
board. All the metallic objects are connected to the ground plane of the board which
covers its bottom plane.

\begin{figure*}[htb!]
\begin{center}
\begin{tabular}{cc}
\includegraphics[height=2.0in,width=1.4in] {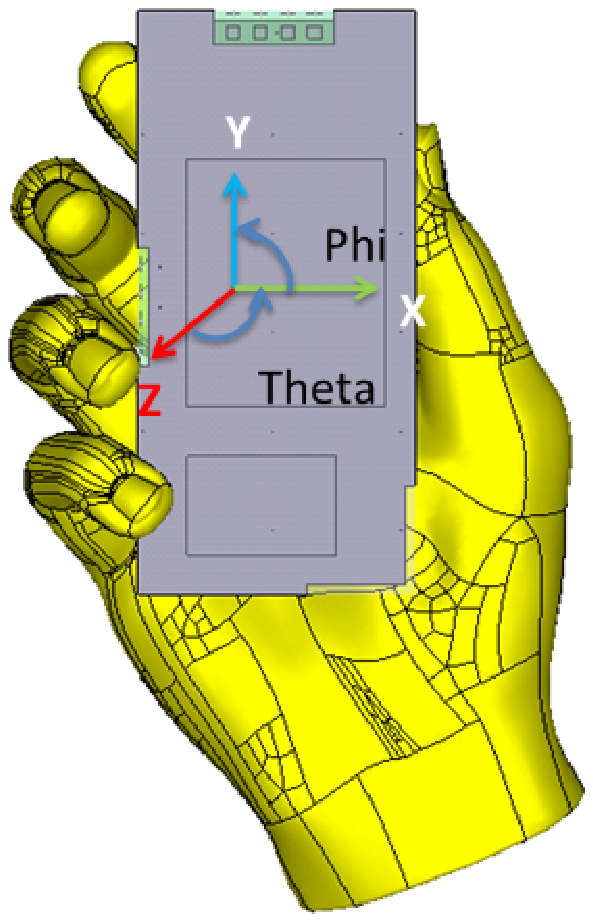}
&
\includegraphics[height=2.0in,width=2.0in] {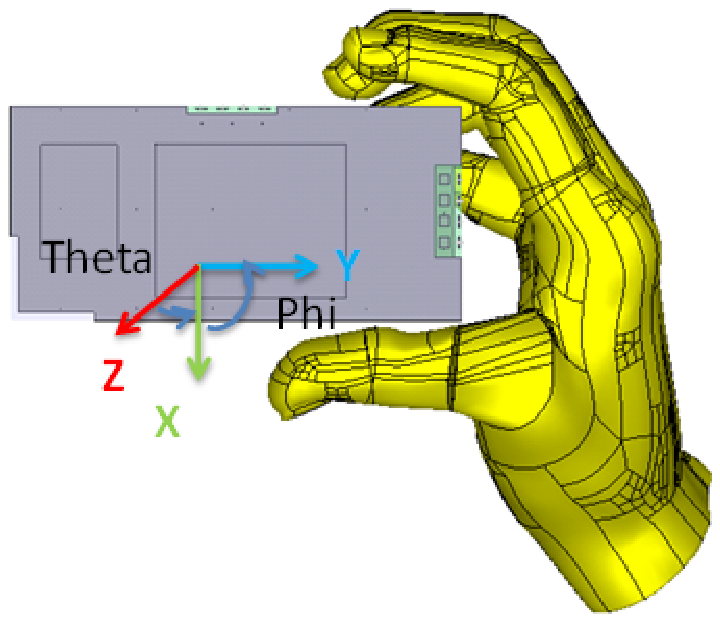}
\\
{\hspace{0.3in}} (a) & (b)
\end{tabular}
\caption{\label{fig_subarray}
A typical UE design with multiple subarrays and a hand phantom model in
(a) {\em Portrait} mode and in
(b) {\em Landscape} mode, along with the local coordinate system capturing
azimuth and elevation angles ($\phi$ and $\theta$).}
\end{center}
\end{figure*}

\begin{figure*}[htb!]
\begin{center}
\begin{tabular}{cc}
\includegraphics[height=1.3in,width=3.0in] {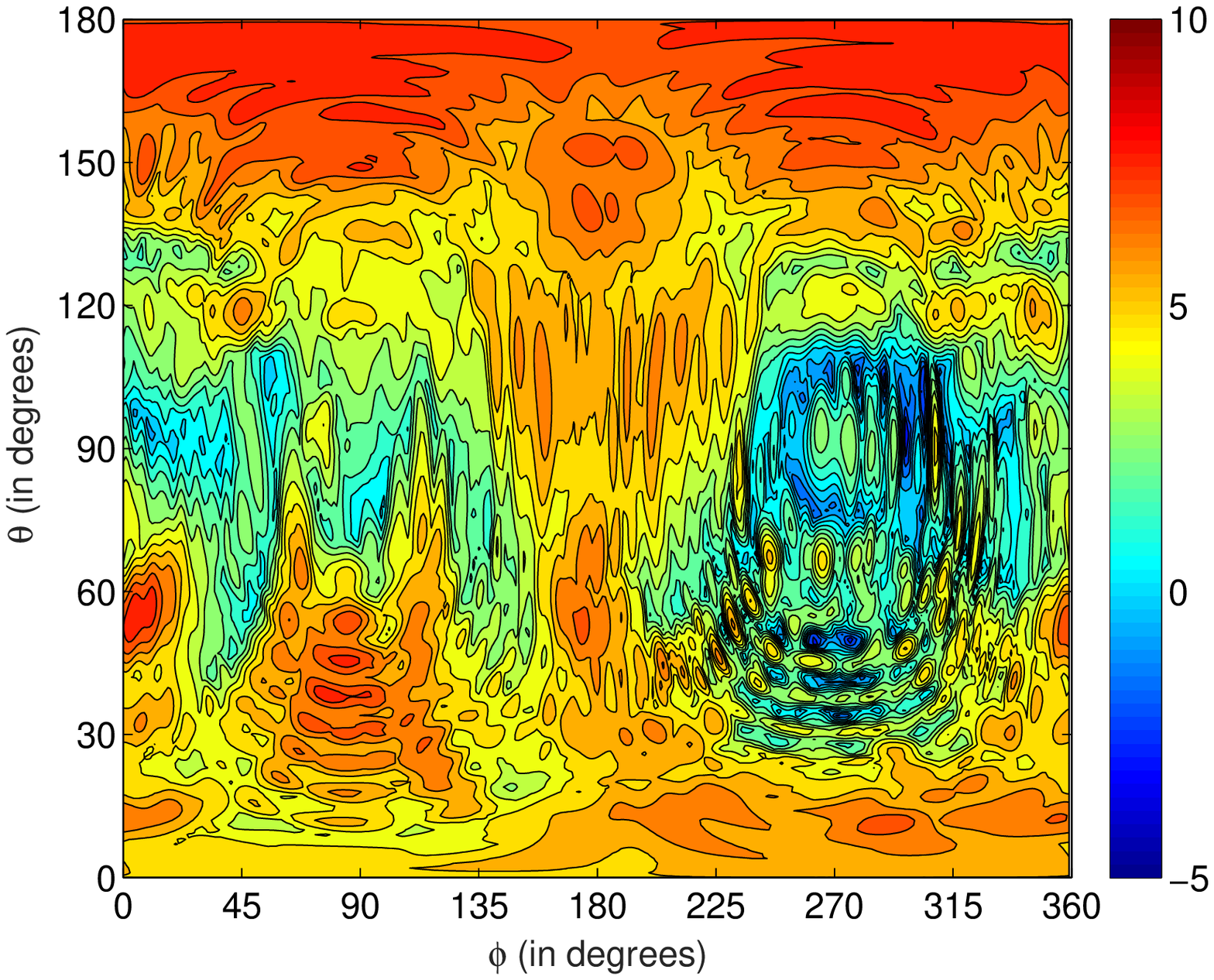}
&
\includegraphics[height=1.3in,width=3.0in] {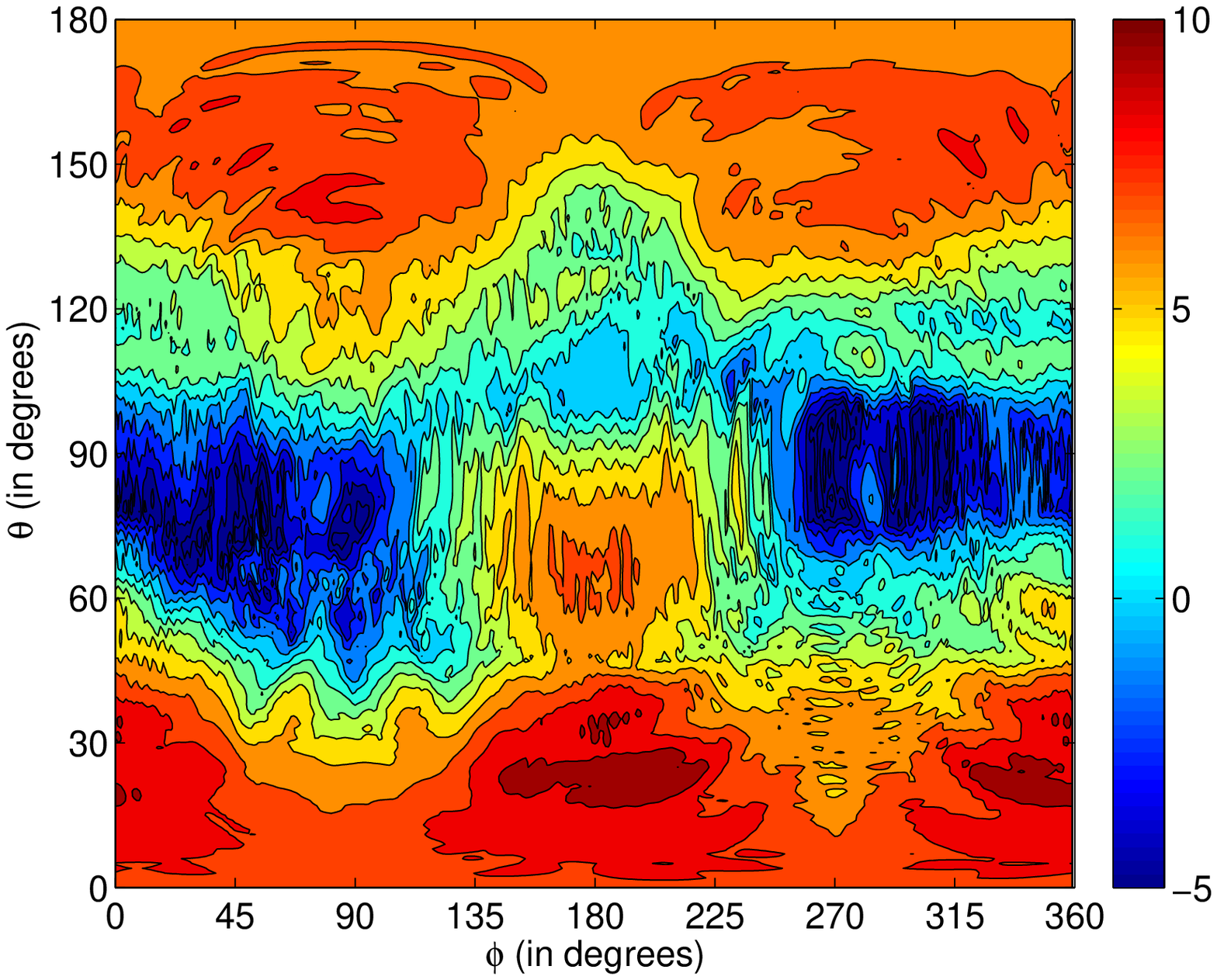}
\\ (a) & (d) \\
\includegraphics[height=1.3in,width=3.0in] {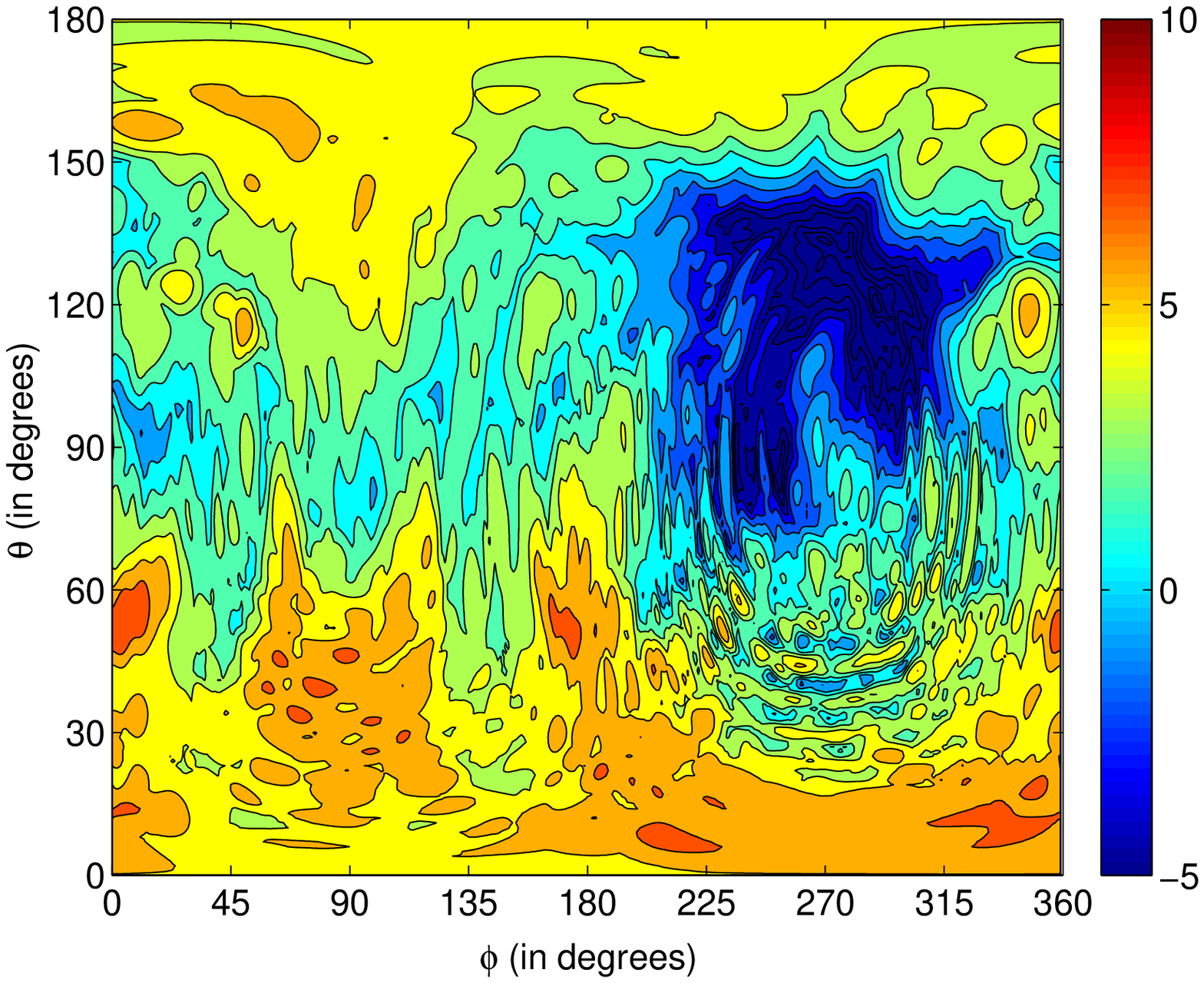}
&
\includegraphics[height=1.3in,width=3.0in] {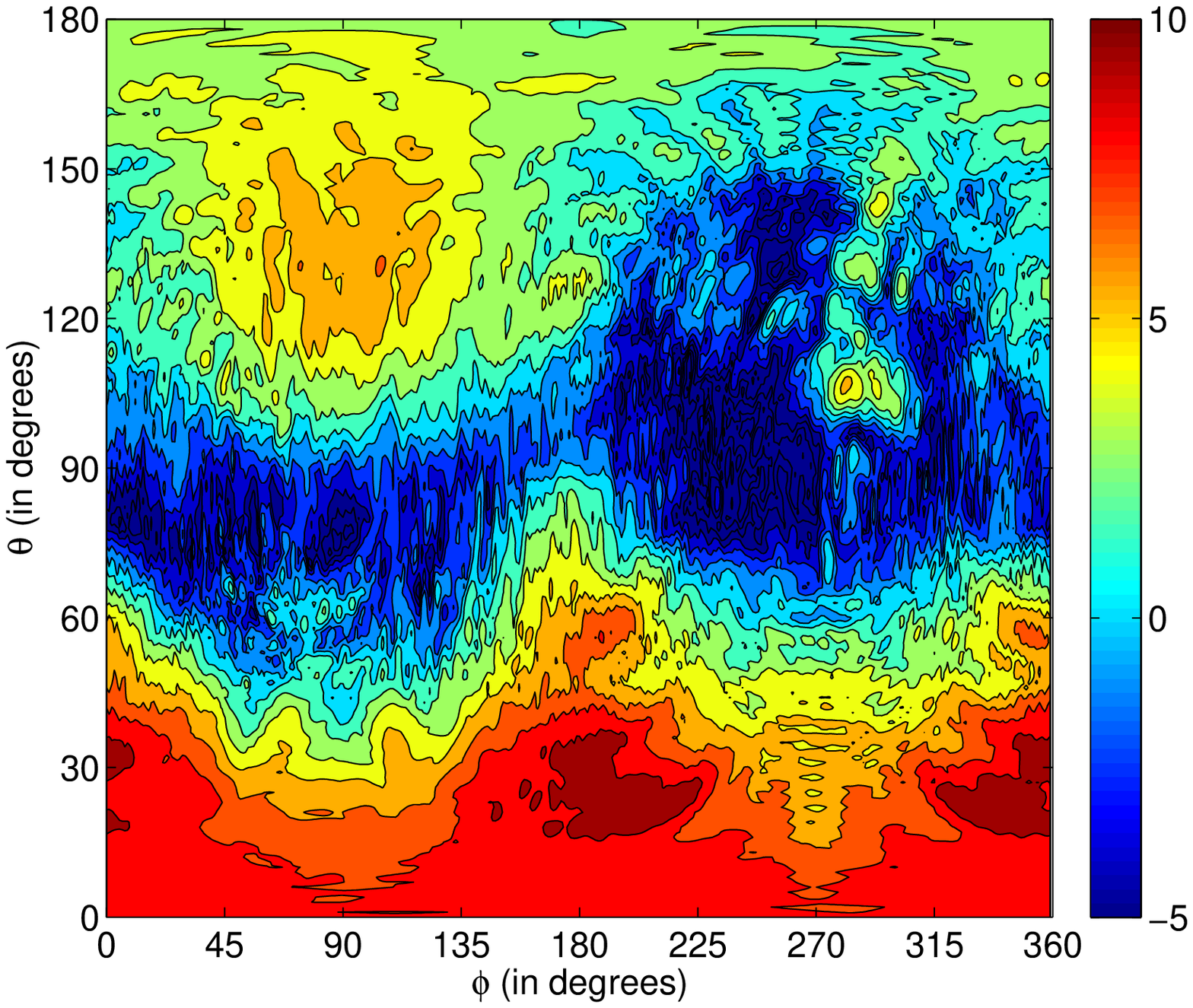}
\\ (b) & (e) \\
\includegraphics[height=1.3in,width=3.0in] {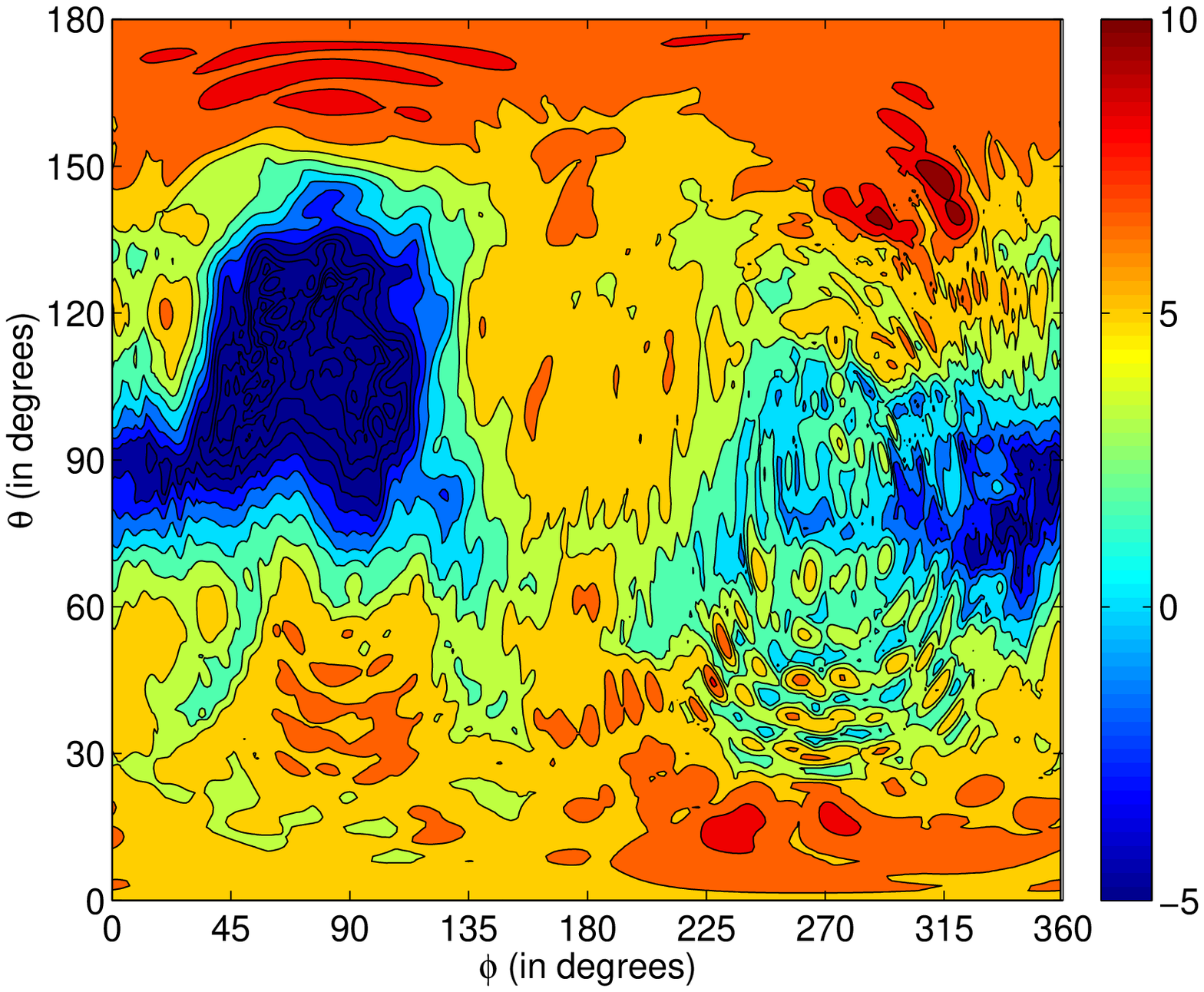}
&
\includegraphics[height=1.3in,width=3.0in] {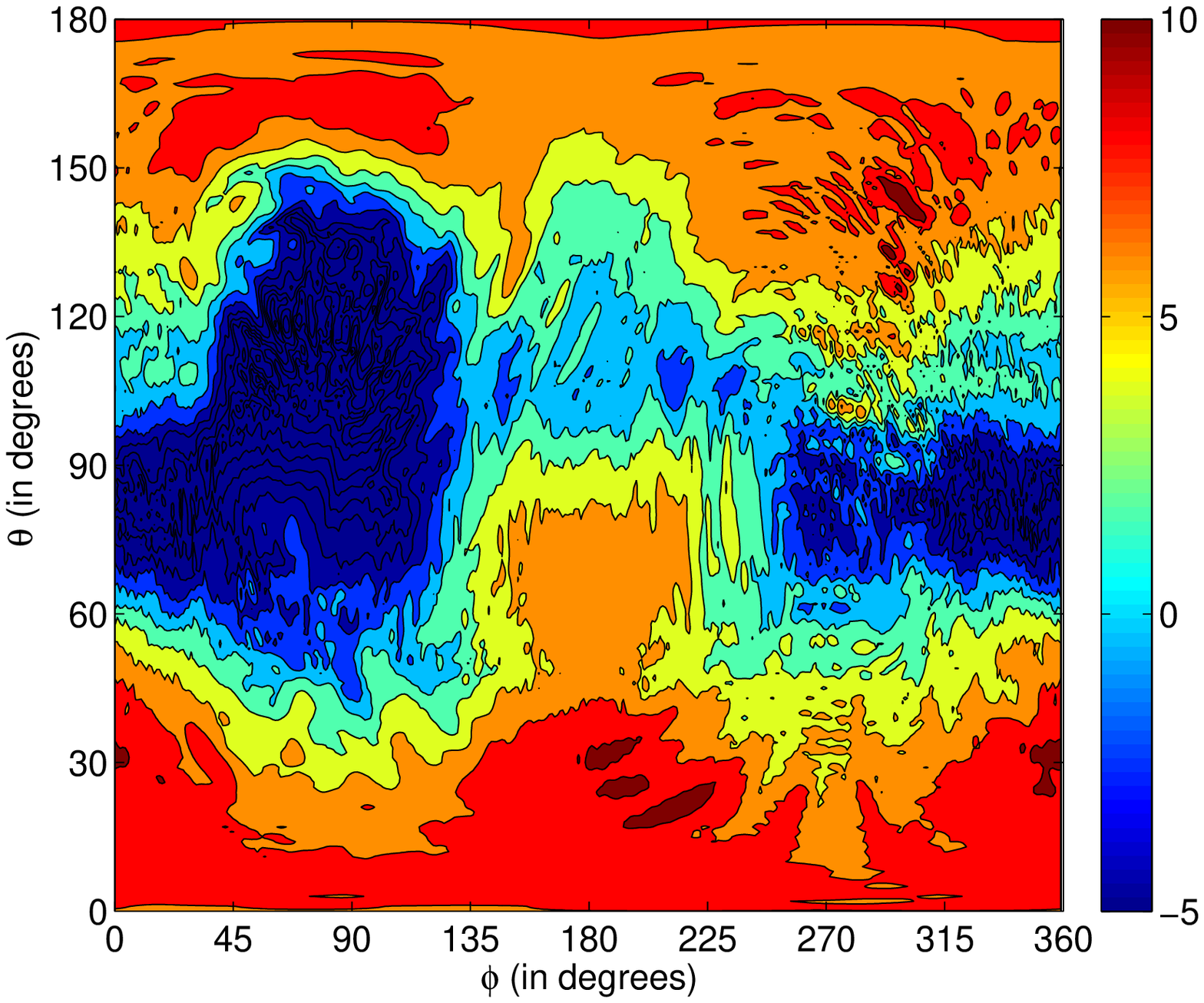}
\\
(c) & (f)
\end{tabular}
\caption{\label{fig_28_60GHz}
Maximum gain of all antennas at $28$ GHz: (a) No hand, (b) Hand in {\em Portrait}
mode, (c) Hand in {\em Landscape} mode; and at $60$ GHz: (d) No hand, (e) Hand in
{\em Portrait} mode, (f) Hand in {\em Landscape} mode.}
\end{center}
\end{figure*}

For the antenna design, multiple subarray units (corresponding to placement of
antennas on the Long and Top edges of the UE) in the {\em Portrait} and in
{\em Landscape} modes, as illustrated in Figs.~\ref{fig_subarray}(a)-(b), are
considered. These
antenna modules are designed on a relatively low loss dielectric substrate (Rogers
4003) and are placed on the FR-4 substrate. The antenna elements are either dipole
elements or dual-polarized patch elements. Antennas are designed for the respective
carrier frequency and are simulated with and without hand at every azimuth and
elevation angle ($\phi$ and $\theta$) in the spherical coordinate system.

To understand the impact of hand, the human hand 
is modelled as a homogeneous dielectric with the dielectric properties of skin tissue.
In general, the skin permittivity decreases with an increase in frequency while its
conductivity increases~\cite{andreuccetti,chahat,gabriel}. In particular, a relative
dielectric constant $\epsilon_r = 16.5$ and conductivity $\sigma = 25.8$ S/m are
used at $28$ GHz with CST Microwave Studio (a commercial electromagnetics simulation
software suite). The hand dielectric properties determine the penetration depth into
the hand and the reflection of electromagnetic waves from the hand. At the range of
frequencies considered ($28$ and $60$ GHz), the penetration depth into the hand is
very small (corresponding to a high degree of reflection) ensuring a high degree of 
blockage by the hand.

The maximum gain of all the antennas without hand, with hand in the {\em Portrait}
mode and with hand in the {\em Landscape} mode are plotted as two-dimensional contour plots
(across $\phi$-$\theta$) in Figs.~\ref{fig_28_60GHz}(a)-(c) for $28$ GHz. From
Fig.~\ref{fig_28_60GHz}(a), in the absence of hand, almost the entire sphere is
covered around the UE illustrating both the necessity and the goodness of the
multi-subarray UE design. On the other hand, the presence of hand in either the
{\em Portrait} or {\em Landscape} modes adversely affects the radiation coverage
as seen from the blue areas\footnote{Note that the heat map adjacent to the gain
plots
illustrate the strength of signal coverage.} (defined as signal strengths below
$-5$ dBi) in Figs.~\ref{fig_28_60GHz}(b)-(c). The region in blue stretches from
behind the palm to the thumb (associated based on the corresponding
$\phi$-$\theta$ angles). Furthermore, it is seen that the long edge module does
not play an important role in signal reception in the {\em Portrait} mode as it is
blocked with the fingers resulting in significantly deteriorated antenna
efficiencies. On the other hand, the short edge is not affected much with the
presence of the hand ensuring that diversity from subarrays is critical for
seamless communications. For the {\em Landscape} mode, the antennas along the long
edge are unobstructed leading to better coverage than the short edge antennas.

While the electrical properties of the hand are different between $28$ and $60$
GHz, the hand still acts as a lossy reflector and can severely deteriorate antenna
performance. In general, the skin permittivity and conductivity decreases and
increases with frequency, respectively. Based on~\cite{andreuccetti,chahat,gabriel},
a relative dielectric constant of $\epsilon_r = 7.9$ and conductivity of
$\sigma = 36.4$ S/m are used at $60$ GHz and simulations (as above) are redone.
Figs.~\ref{fig_28_60GHz}(d)-(f) plot the coverages without hand, with hand in the
{\em Portrait} and {\em Landscape} modes, respectively, for $60$ GHz. Compared to $28$ GHz, at
$60$ GHz, the electrical size of the UE and hand are almost doubled while the
antenna size is approximately reduced to half. This prevents the radiation of the
antennas located on the top and left sides to leak to the right or bottom sides
of the UE. Thus, even in the absence of hand, complete radiation coverage around
the UE is quite challenging (as seen from the streaks of blue without the hand in
Fig.~\ref{fig_28_60GHz}(d)).

In the {\em Portrait} mode in Fig.~\ref{fig_28_60GHz}(e), the
left edge of the phone is obstructed by the fingers which causes blockage on the
left side as well as back side of the hand. This forms a dead zone between
$60^{\sf o}$ to $110^{\sf o}$ from the Z-axis, which is perpendicular to the hand.
The high gain radiation coverage occurs in front of the phone due to radiation
from the patch antennas located at the top edge. In comparison, the back of the
phone which is illuminated by the top edge dipoles shows slightly lower gain. In
the {\em Landscape} mode in Fig.~\ref{fig_28_60GHz}(f), while neither subarray at the UE
is blocked severely, the hand still prevents the top edge dipoles to radiate towards
the back of the hand and most of the radiation is reflected or absorbed.

\subsection{Loss in Signal Strength}
\label{sec2b}
\begin{figure*}[htb!]
\begin{center}
\begin{tabular}{cc}
\includegraphics[height=2.3in,width=2.9in]
{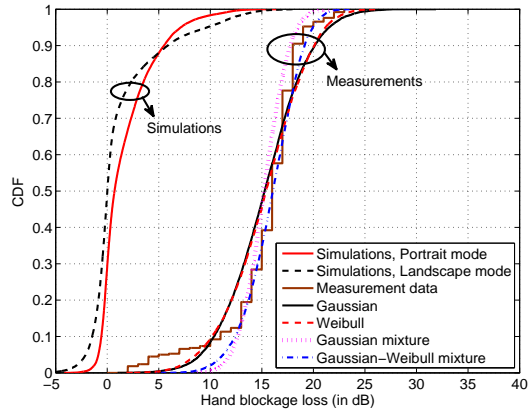}
&
{\hspace{0.5in}}
\includegraphics[height=2.2in,width=1.4in] {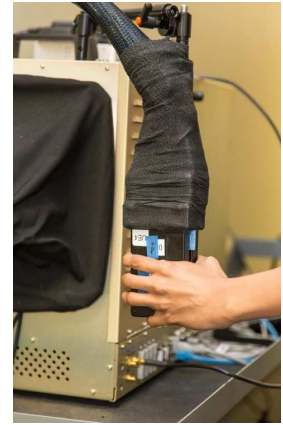}
\\ (a) & {\hspace{0.5in}} (b)
\end{tabular}
\caption{\label{fig_handloss}
(a) CDF of hand blockage loss with electromagnetic simulations using a hand phantom
model, measurements using the experimental prototype, and model fits to measurement
data. (b) Pictorial illustration of the hand holding experiment with the UE where
the hand covers/envelops the active antenna array completely.}
\end{center}
\end{figure*}

The blockage loss incurred by the hand in the studies of Sec.~\ref{sec2a} 
are plotted in Fig.~\ref{fig_handloss}(a) [see curves grouped as ``Simulations'']. In
particular, the cumulative distribution function (CDF) of the maximal gain from all
the antennas in both polarizations is compared between the {\em Freespace} mode
(without the hand) and {\em Portrait}/{\em Landscape} modes (with hand) and plotted here. These
studies show that the loss appears to be in the range of $-5$ to $15$ dB (with negative
values corresponding to signal energy boosting due to hand reflection at certain angles).
In particular, Fig.~\ref{fig_handloss}(a) shows that blockage loss is possible for up
to $70\%$ and $50\%$ of the angles in {\em Portrait} and {\em Landscape} modes, respectively.
However, these estimates are sensitive to the 
hand phantom model used in the simulation studies (see 
Fig.~\ref{fig_subarray}). In particular, the tightness of the hand grip around
the UE determines what fraction of electromagnetic energy is captured by the antennas
with a tighter/harder grip leading to significant losses (and {\em vice versa}). The
nature of the air gap between the fingers also determines what fraction of energy is
captured as multiple reflections from the skin surface, which can assist signal
reception in certain angles. Thus, while the above simulation studies could be used
for studying loss in angular coverage, using them directly for loss in signal strength
could be questionable.

In this context, some measurement studies
from~\cite{maccartney_vtc2016,maccartney_2017,5G_whitepaper} suggest a loss of
even up to $30$ to $40$ dB with hand and body blockage. However, all these
measurements are based on the use of horn antennas and cannot be extended easily to
form-factor UE designs. To understand the loss possible with the hand, measurement
studies are performed with a $28$ GHz experimental prototype capturing the
attributes of a 5G base-station as well as a form-factor UE design and
operating in a time-division duplexing framework. 
With this prototype, the baseband analog in-phase and quadrature (IQ) signals
are routed to/from the modem to an IQ modulator/demodulator at $2.75$ GHz center
frequency. The $2.75$ GHz intermediate frequency signal is translated to $28$ GHz
using a $25.25$ GHz tunable local oscillator (with a $100$ MHz step size). The
base-station end of this system is a $16 \times 8$ planar array (made of a
waveguide design) that allows analog beamforming using tunable four bit phase
shifters and gain controllers. The UE end is a form-factor design made of four
selectable subarrays, each a four element phased array of either dipoles or
patches with locations on the UE as illustrated in Fig.~\ref{fig_subarray}(a).
More details on the design parameters of the prototype are provided
in~\cite{vasanth_comm_mag_16}.

In our hand loss studies, the form-factor UE is grabbed by the hand at normal speed
and the hand completely covers/envelops the active antenna arrays as illustrated in
Fig.~\ref{fig_handloss}(b). All the subarrays at the UE side except the enveloped
subarray are disabled in terms of beam switching thus allowing us to capture the
blockage loss in terms of received signal strength differentials. We define an
{\em RF event}
relevant for these studies as one where the received signal strength indicator
(RSSI) drops from the steady-state value by at least $2$ dB. From our experiments,
$38$ such RF events are recorded and for each RF event, ten RSSI minimas\footnote{Due
to constraints on the post-processing of data, RSSI minimas can be recorded to only
within a $1$ dB precision.} spanning the entire event are separately recorded. The link
degradation is computed as the RSSI difference between the steady-state RSSI value
and the ten minimas. The empirical CDF of hand blockage loss corresponding to these
$380$ data points is also plotted in Fig.~\ref{fig_handloss}(a) [grouped under
``Measurements'']. From this plot, we note that observations of $30$-$40$ dB hand
blockage loss
from~\cite{maccartney_vtc2016,maccartney_2017,5G_whitepaper} may be too pessimistic
to capture real measurements based on form-factor UE designs. More reasonable estimates
for hand blockage loss are on the order of $5$-$20$ dB with a median loss of $15$ dB.

We now consider a number of model fits for the hand blockage loss data. The efficacy
of each model (for the loss data) is captured by the Kolmogorov-Smirnov (KS)
distance~\cite{subha}, defined as,
\begin{eqnarray}
d_{ {\sf KS}}( {\sf data}, {\sf model})
\triangleq \max_x \big|F_{\sf data}(x) - F_{\sf model}(x) \big|,
\nonumber
\end{eqnarray}
where $F_{\sf data}(x)$ and $F_{\sf model}(x)$ denote the CDFs with the empirical
data and the corresponding model, respectively. Since the KS distance only
captures the worst-case
deviation between the two CDFs, we also consider a data-{\em weighted} KS (WKS)
distance, defined as,
\begin{eqnarray}
d_{ {\sf WKS}}( {\sf data}, {\sf model})
\triangleq \int x \cdot \big|F_{\sf data}(x) - F_{\sf model}(x) \big| \cdot dx.
\nonumber
\end{eqnarray}
A better model fit for the data is captured by smaller KS and WKS distances (and
{\em vice versa}).

\begin{table*}[htb!]
\caption{Model Fits for Hand and Body Blockage Loss} 
\label{table_blockage_loss_models}
\begin{center}
\begin{tabular}{|c||c|c|c||c|c|c|}
\hline
&
\multicolumn{3}{c||}{Hand blockage} &
\multicolumn{3}{c|}{Body blockage}
\\ \hline
Model & Parameters & $d_{\sf KS}$ & $d_{\sf WKS}$
& Parameters & $d_{\sf KS}$ & $d_{\sf WKS}$
\\
\hline
\hline
Gaussian: $\frac{1} {\sqrt{2 \pi \sigma^2} } \cdot
e^{ - \frac{ (x-\mu)^2}{2 \sigma^2} }$
& $\mu = 15.26$, $\sigma = 3.80$ & $0.19$ & $1.22$ &
$\mu = 8.54$, $\sigma = 2.45$ & $0.17$ & $0.46$
\\ \hline
Weibull: $\frac{\beta}{\alpha} \cdot \left( \frac{x}{\alpha} \right)^{\beta - 1}
\cdot 
e^{ - \left( \frac{x}{\alpha} \right)^{\beta} }$
& $\alpha = 16.70$, $\beta = 4.61$ & $0.17$ & $1.16$
& $\alpha = 9.43$, $\beta = 3.94$ & $0.16$ & $0.45$
\\ \hline
Gaussian mixture: &
$p_1 = 0.75$, $\mu_1 = 16.28$, & $0.26$ & $1.51$ &
$p_1 = 0.11$, $\mu_1 = 3.23$, & $0.21$ & $0.62$
\\ 
$ \frac{ p_1} {\sqrt{2 \pi \sigma_1^2} }
\cdot e^{ - \frac{ (x-\mu_1)^2}{2 \sigma_1^2} } +$ &
$\sigma_1 = 1.71$, $p_2 = 0.25$, & & &
$\sigma_1 = 0.42$, $p_2 = 0.89$, & &
\\
$ \frac{ 1 - p_1} {\sqrt{2 \pi \sigma_2^2} }
\cdot e^{ - \frac{ (x-\mu_2)^2}{2 \sigma_2^2} } $ &
$\mu_2 = 12.15$, $\sigma_2 = 6.03$ & & &
$\mu_2 = 9.17$, $\sigma_2 = 1.70$ & & \\
\hline
Gaussian Weibull mixture: &
$p_1 = 0.15$, $\mu = 15.76$, & $0.14$ & $0.70$ &
$p_1 = 0.15$, $\mu = 9.54$, & $0.15$ & $0.39$
\\
$\frac{p_1} {\sqrt{2 \pi \sigma^2} } \cdot
e^{ - \frac{ (x-\mu)^2}{2 \sigma^2} } +$ &
$\sigma = 3.55$, $p_2 = 0.85$, & & &
$\sigma = 1.95$, $p_2 = 0.85$, & &
\\
$\frac{(1 - p_1) \cdot \beta}{\alpha} \cdot \left( \frac{x}{\alpha} \right)^{\beta - 1}
\cdot 
e^{ - \left( \frac{x}{\alpha} \right)^{\beta} }$ &
$\alpha = 17.20$, $\beta = 6.11$ & & &
$\alpha = 9.43$, $\beta = 3.69$ & & \\
\hline \hline
\end{tabular}
\end{center}
\end{table*}

The empirical mean and empirical standard deviation of the hand blockage loss
data are $15.26$ dB and $3.80$ dB, respectively. The functional forms as well
KS and WKS distances of the different models considered here are presented in
Table~\ref{table_blockage_loss_models}. In terms of the models, we first consider
a Gaussian density\footnote{Theoretically, a Gaussian approximation to the loss
can result in a value that is negative due to the Gaussian tail. Independent of
whether negative loss values make sense, for all practical purposes, such
realizations are not seen in numerical studies due to the extremely low probability
of such occurrences. Thus, we will not bother with these technical difficulties in
this paper.} with mean and standard deviation being the
empirical mean and empirical standard deviation values (as above). This model fit
is plotted in 
Fig.~\ref{fig_handloss}(a). From this plot, we observe that the Gaussian density
over-estimates both the lower and upper tail values of loss and under-estimates
the bulk. That is, the measured data has a heavier lower
tail than the Gaussian density can fit. This mismatch is also reflected in the high
KS and WKS distances with this model (see Table~\ref{table_blockage_loss_models}).
To overcome this problem, we consider a Weibull density\footnote{Note that the
Weibull density $W (\alpha, \beta)$, where $\alpha$ and $\beta$ are the scale and
shape parameters, is commonly used to model ``time-to-failure'' of a certain process
with $\beta$ capturing the failure rate of the process. In particular, $\beta < 1$,
$\beta = 1$ and $\beta > 1$ capture decreasing, constant, and increasing failure
rates of the process with time.} that is typically used to model heavy tailed data.
While the Weibull density improves the KS and WKS distances, from
Fig.~\ref{fig_handloss}(a) as well as
Table~\ref{table_blockage_loss_models}, 
we see that there is no dramatic improvement in fit with this model. Note that both
these models are described by two parameters.

For a better fit, we therefore consider models with more parameters. While different
model families can be considered, we start with a Gaussian mixture with the hope that
these Gaussians can individually capture the upper and lower tails better. The
parameters for the Gaussian mixture are learned using the Expectation Maximization
algorithm as described in Appendix~\ref{app_model_GMM}. In contrast to the expectation,
our results show that the best Gaussian mixture model fails to capture the data
accurately as the tails of one Gaussian component lead to a poor fit for the other
Gaussian component (and {\em vice versa}). In fact, the Gaussian mixture leads to a
good fit only at the bulk. To remedy this misfit, we consider a Gaussian-Weibull
mixture model as an alternate candidate. Since an analogous Expectation Maximization
solution (as in Appendix~\ref{app_model_GMM}) appears to be difficult due to the
complicated structure of the Weibull density,
we use a local search over the parameter space neighborhood initialized with the
individual Gaussian model and Weibull model parameters, respectively. The objective of
this optimization is to minimize the WKS distance of the consequent model fit. As
observed from both Fig.~\ref{fig_handloss}(a) and Table~\ref{table_blockage_loss_models},
the optimized Gaussian-Weibull mixture fits the empirical data better suggesting its
utility as a generative model for hand blockage loss.

\subsection{Proposed Statistical Model}
\label{sec2c}
Based on the studies in Sec.~\ref{sec2a} and~\ref{sec2b}, we propose a
simple square region approximation for spatial/angular blockage with the
hand. This approximation is captured by the center of the blocker ($\phi_1$,
$\theta_1$), and the angular spread of the blocker ($x_1$, $y_1$) in azimuth
and elevation with the blocking angles captured as $\phi \in \left[ \phi_1 -
\frac{x_1}{2}, \hspp \phi_1 + \frac{x_1}{2} \right]$ and $\theta \in
\left[ \theta_1 - \frac{y_1}{2}, \hspp \theta_1 + \frac{y_1}{2} \right]$ in
azimuth and elevation, respectively. For the blockage loss, we propose a
low-complexity variant captured by the Gaussian fit and a relatively
higher-complexity variant captured by the Gaussian-Weibull mixture fit from
Table~\ref{table_blockage_loss_models}. These proposals lead to the
statistical model in Table~\ref{table_proposed_model_selfblockage} (in
the local coordinate system around the UE) for self-blockage.

\begin{table*}[htb!]
\caption{Proposed Statistical Model for Self-Blockage ($k = 1$)}
\label{table_proposed_model_selfblockage}
\begin{center}
\begin{tabular}{|c||c|c|c|c|c|}
\hline
Scenario & $\phi_1$ & $x_1$ & $\theta_1$ & $y_1$ & Blockage loss (in dB) \\
\hline
\hline
{\em Portrait} mode & $260^{\sf o}$ & $120^{\sf o}$ & $100^{\sf o}$ &
$80^{\sf o}$ & 
Low-complexity\hspp: \hspp
${\cal N}(\mu = 15.3 \hspp {\sf dB}, \hspp \sigma = 3.8 \hspp {\sf dB})$
\\
\cline{1-5}
{\em Landscape} mode & $40^{\sf o}$ & $160^{\sf o}$ & $110^{\sf o}$ &
$75^{\sf o}$ & 
High-complexity\hspp: \hspp
Gaussian-Weibull mixture with
\\
& & & & & $p_1 = 0.15$, $\mu = 15.8$, $\sigma = 3.6$,
$p_2 = 0.85$, $\alpha = 17.2$, and $\beta = 6.1$
\\
\hline
\hline
\end{tabular}
\end{center}
\end{table*}

\section{Dynamic Blockage}
\label{sec3}

\subsection{Methodology for Modeling Dynamic Blockage} 
\label{sec3a}
We assume that the dominant signal path(s) between the transmitter (base-station)
and the receiver
(UE) are in the plane connecting them. In outdoor use-cases where the transmitter
is on the top of a building or a lamp post, as well as indoor use-cases where
the transmitter is on/near the ceiling or at the same level as the receiver, such
an assumption is reasonable\footnote{Nevertheless, this assumption significantly
simplifies the study done in this work and should be treated as a first attempt
at a comprehensive statistical model for blockage. Future studies will consider
further extensions of the setup considered in this work.} with at least moderate
cell sizes. We now propose a simple methodology to 
capture the impact of dynamic blockers (e.g., humans, vehicles, etc.)
that are strewn randomly in the transmit-receive plane.

Relative to a global coordinate system, the azimuth angle of the blockers are
assumed to be uniform in $[0^{\sf o}, \hsppp 360^{\sf o})$ and the elevation
angle is assumed to be a fixed $\theta_{\sf o}$. Without loss in generality,
we assume that $\theta_{\sf o} = 90^{\sf o}$ (that is, the transmit-receive
plane is the horizontal plane). If the blockers are too close to either the
transmitter or the receiver, the observed blockage loss can be significant.
At the receiver end, such a scenario is already captured by the self-blockage
component and at the transmitter end, it is less likely provided the transmitter
has a reasonable unobstructed coverage (the precise scenarios considered in this
work).

\begin{figure*}[htb!]
\begin{center}
\begin{tabular}{cc}
\includegraphics[scale = 0.6] {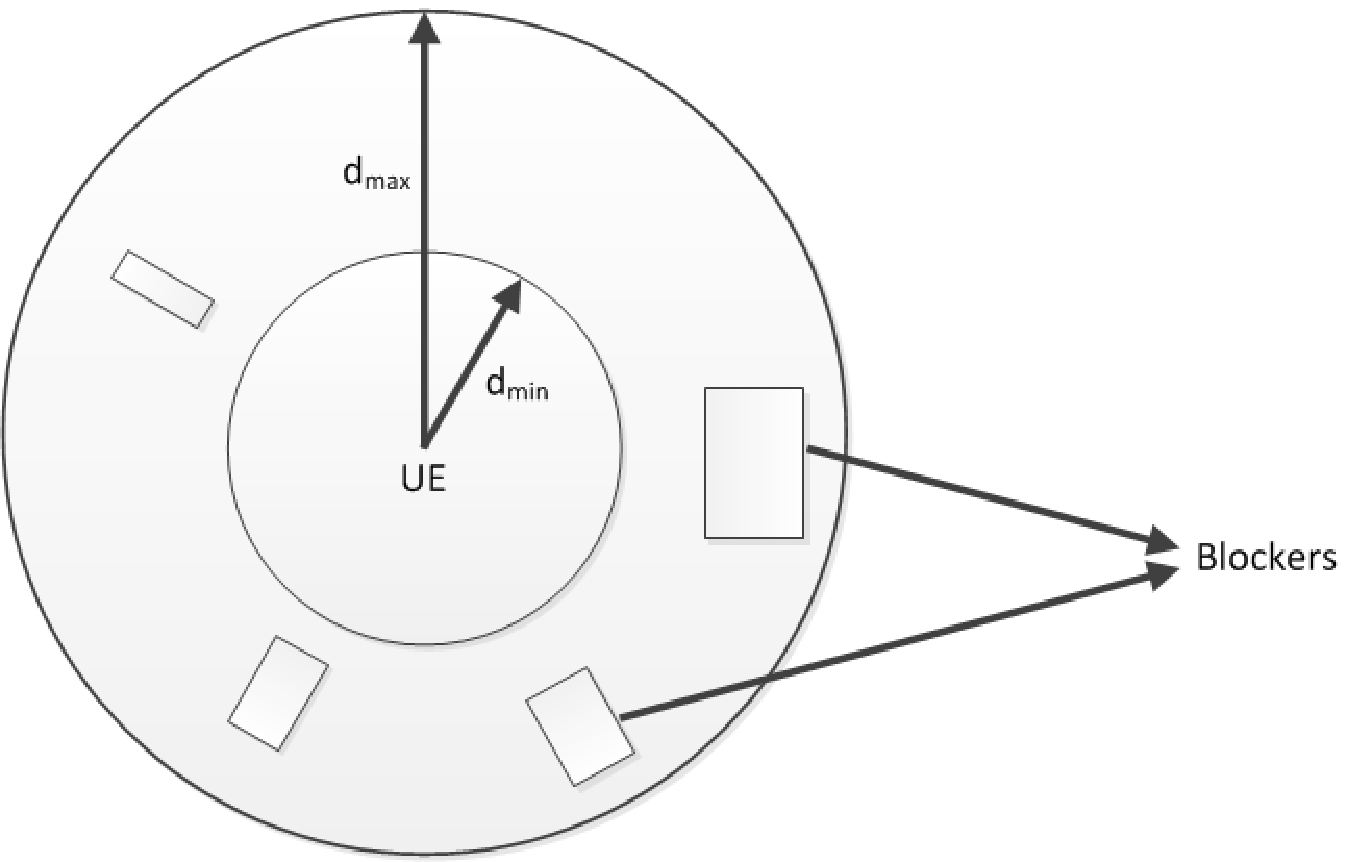}
&
\includegraphics[height=2.2in,width=1.4in] {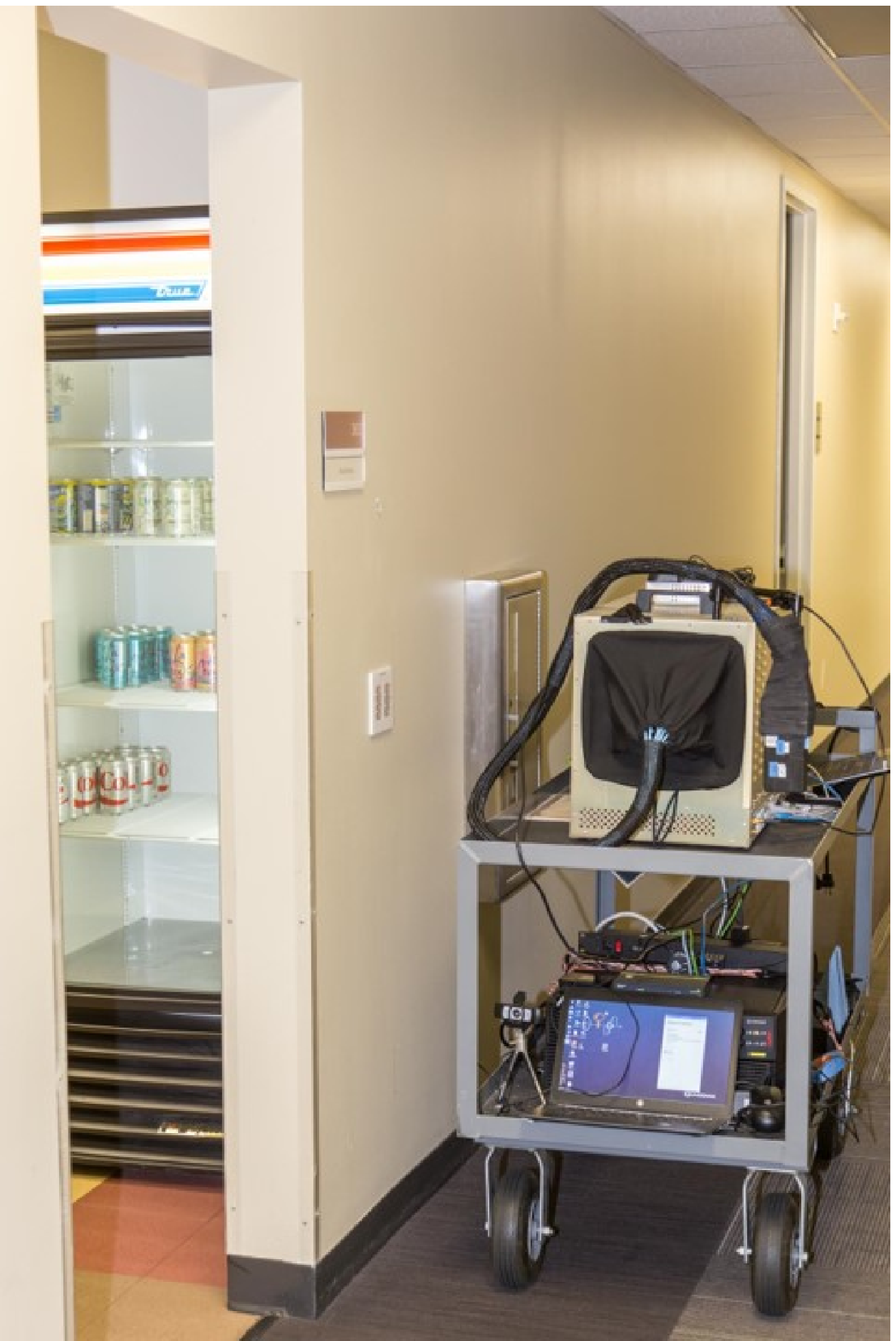}
\\
\\
{\hspace{-1.3in}}
(a) & (b)
\end{tabular}
\caption{\label{fig_scenario_humans_vehicles}
(a) Top-view of simulation setup for studying dynamic
blockage modeling. (b) Measurement scenario for body blockage studies.}
\end{center}
\end{figure*}

To capture these aspects explicitly, as illustrated in the top-view of the
transmit-receive plane in Fig.~\ref{fig_scenario_humans_vehicles}(a), the
blockers are assumed to lie in a circular region (with radial locations $r$
constrained as $d_{\sf min} \leq r \leq d_{\sf max}$) around the receiver. A
triangular density function
of the form
\begin{eqnarray}
f(r) = \frac{ 2 \left( r - d_{\sf min} \right) }
{ \left( d_{\sf max} - d_{\sf min} \right)^2 },
\hspp \hspp d_{\sf min} \leq r \leq d_{\sf max}
\nonumber
\end{eqnarray}
is assumed for $r$. This density captures the fact that the blocker density
grows with $r$ since there is more area covered by the circular region with
$r$ closer to $d_{\sf max}$ than at $d_{\sf min}$.
The heights and widths of the blockers are modeled as 
$h \sim
{\cal U} \left( [\overline{\sf H} - {\sf h}, \hsppp \overline{\sf H} +
{\sf h} ] \right)$ and $w \sim 
{\cal U} \left( [\overline{\sf W} - {\sf w}, \hsppp
\overline{\sf W} + {\sf w} ] \right)$
where ${\cal U}\left( [a , b] \right)$ stands for a uniform random variable
over the interval $[a, b]$, $\overline{\sf H}$ and $\overline{\sf W}$ denote
the mean height and width of the blocker, and ${\sf h}$ and ${\sf w}$ denote
the one-sided deviations for the height and width of the blocker, respectively.

We propose to model the number of blockers in the region of interest as a Poisson
random variable with parameter $\lambda$. That is,
\begin{eqnarray}
{\sf P}\left( {\sf No. \hspp of \hspp blockers} = k\right) = \frac{\lambda^k}{k!} \cdot
e^{-\lambda}, \hspp k = 0, 1, 2, \cdots
\nonumber
\end{eqnarray}
Let the {\em average density of blockers}
be defined as the number of blockers per unit area. With the model from
Fig.~\ref{fig_scenario_humans_vehicles}(a), 
the average density of blockers is given as
$\frac{ \lambda} { \pi \left(  d_{\sf max}^2 - d_{\sf min}^2 \right) }$. For
simplicity, the blocker and the receiver are assumed to be parallel in orientation
and the angle subtended at the receiver/UE by the blocker is
computed\footnote{Since the distances tend to be small for both indoor and
outdoor use-cases, approximations such as $\phi = \frac{w}{r}$ and
$\theta = \frac{h}{r}$ can be inaccurate.} as
$\sin \left( \frac{\phi}{2} \right) = \frac{ w/2}{r}$ and
$\sin \left( \frac{\theta}{2} \right) = \frac{ h/2} {r}$ in
azimuth and elevation, respectively.

\begin{table*}[htb!]
\caption{Average Density of Blockers with Different $d_{\sf max}$
and $d_{\sf min}$}
\label{table_average_density}
\begin{center}
\begin{tabular}{|c|c|c|c||c|c|c|c|}
\hline
Human ($d_{\sf min} = 3$ m)
& $\lambda = 4$ & $\lambda = 8$ & $\lambda = 12$ &
Vehicular ($d_{\sf min} = 5$ m) &
$\lambda = 4$ & $\lambda = 8$ & $\lambda = 12$ \\
\hline
$d_{\sf max} = 10$ m & $0.0140$ & $0.0280$ & $0.0420$ &
$d_{\sf max} = 30$ m & $0.0015$ & $0.0029$ & $0.0044$ \\
\hline
$d_{\sf max} = 15$ m & $0.0059$ & $0.0118$ & $0.0177$ &
$d_{\sf max} = 40$ m & $0.0008$ & $0.0016$ & $0.0024$ \\
\hline
$d_{\sf max} = 20$ m & $0.0033$ & $0.0065$ & $0.0098$ &
$d_{\sf max} = 50$ m & $0.0005$ & $0.0010$ & $0.0015$ \\
\hline
\hline
\end{tabular}
\end{center}
\end{table*}

To study the loss in spatial/angular coverage, we use the parameters
from~\cite{5G_whitepaper}: $\overline{\sf H} = 1.7$ m and $\overline{\sf W}
= 0.3$ m for human blockers, and  $\overline{\sf H} = 1.4$ m and
$\overline{\sf W} = 4.8$ m for vehicular blockers. For modeling human and
vehicular variations, we also use ${\sf h} = 0.2$ m and ${\sf w} = 0.1$ m
for humans, and ${\sf h} = 0.4$ m and ${\sf w} = 0.5$ m for vehicles.
In a typical indoor office setting such as the third floor of the Qualcomm
building, Bridgewater, NJ with dimensions of $75 \times 40$ sq m
and occupied by $50$ to $100$ humans, the average density of human blockers
ranges from $0.0166$ to $0.0333$ per sq m. Similarly, in a typical outdoor
setting of $100 \times 100$ sq m with $10$ to $50$ vehicles, the average
density of vehicular blockers could range from $0.001$ to $0.005$ per sq m.
For three choices of $\lambda$ and $d_{\sf max}$, Table~\ref{table_average_density}
presents the average density of human and vehicular blockers in the indoor and
outdoor cases, respectively. For a human blocker, we assume $d_{\sf min} = 3$ m
and three cases for $d_{\sf max}$: $d_{\sf max} = 10$, $15$, or $20$ m (denoted
as Cases $1$-$3$). For a vehicular blocker, we consider $d_{\sf min}
= 5$ m and three cases for $d_{\sf max}$: $d_{\sf max} = 30$, $40$, or $50$ m
(also denoted as Cases $1$-$3$). While different choices of $\lambda$
can be considered in general, from Table~\ref{table_average_density}, we focus
on $\lambda = 4, \hsppp 8, \hsppp 12$ as representative samples to reflect
average density of blockers in typical indoor and outdoor deployments.

\subsection{Loss in Spatial/Angular Coverage and Number of Blockers}
\label{sec3b}
Table~\ref{table_mean_blockage_other_blockers} presents the median, $90^{\sf th}$
and $95^{\sf th}$ percentile values of the mean angular blockage in azimuth
and elevation for $\lambda = 4, 8$ and $12$. From these studies, we note that the mean
spatial/angular blockage for both human and vehicular blockers decreases as
$d_{\sf min}$ increases or $d_{\sf max}$ increases. This is because the blockers
are farther away from the receiver as $d_{\sf min}$ increases and are more likely
to be farther away from the receiver as $d_{\sf max}$ increases. In either case,
a smaller angle is casted at the receiver leading to a reduction in the mean
angular blockage. Further, this angle has only a weak dependence on $\lambda$.
We conclude that a typical mean angular blockage of $2.5^{\sf o}$ in azimuth
and $15^{\sf o}$ in elevation are seen with human blockers, and $15^{\sf o}$ in
azimuth and $5^{\sf o}$ in elevation are seen for vehicular blockers.

We are now interested in understanding the number of blockers $K$ to be incorporated
in a statistical model for dynamic blockage. Table~\ref{table_topK} presents the
explanatory power captured by the fraction of the total azimuthal angular blockage
captured\footnote{While we consider
the top-$K$ blockers in terms of angles blocked, alternate criteria such as top-$K$
blockers in terms of blockage loss can also be considered. The flavor of
the results are not expected to change with such alternate criterion.} by the top-$K$
blockers. The median, $90^{\sf th}$ and $95^{\sf th}$ percentile values of the
explanatory power are presented for human blockers with $d_{\sf min} = 3$ m, $d_{\sf max}
= 15$ m, and for vehicular blockers with $d_{\sf min} = 5$ m, $d_{\sf max} = 40$ m for
different choices of $\lambda$. Table~\ref{table_topK} shows that there is a decreasing
explanatory power for a fixed $K$ as $\lambda$ increases, and diminishing returns
and increasing model complexity with an increase in $K$ for any $\lambda$. The use
of the top-$4$ human and top-$3$ vehicular blockers can explain over $60 \%$ of the
blocked angular region up to $\lambda = 8$ and suggests a good tradeoff point/compromise
between explanatory power and model complexity.

\begin{table*}[htb!]
\caption{Angular blockage metrics with human and vehicular blockers}
\label{table_mean_blockage_other_blockers}
\begin{center}
\begin{tabular}{
|l|c|  c|c|c||  c|c|c|| c|c|c|| c|c|c|} 
\hline
& &
\multicolumn{6}{c||} { 
Mean angular blockage (Human) 
}
& 
\multicolumn{6}{c|} { 
Mean angular blockage (Vehicular) 
}
\\ \hline
& & \multicolumn{3}{c||}{ 
Azimuth (in degrees) 
}
& \multicolumn{3}{c||} {  
Elevation (in degrees) 
}
& \multicolumn{3}{c||}{ 
Azimuth (in degrees) 
}
& \multicolumn{3}{c|} {  
Elevation (in degrees) 
}
\\ \hline
& 
Percentiles 
&
$50$ & $90$ & $95$ &
$50$ & $90$ & $95$ &
$50$ & $90$ & $95$ &
$50$ & $90$ & $95$
\\ \hline
\multirow{3}[6]{0.2cm}{\rotatebox[origin=c]{90}{{\hspace{0.2in}}$\lambda=4$}}
& Case 1 
& $2.34$ & 
$3.00$ & $3.26$
& $13.19$ & 
$16.34$ & $17.55$
& $14.27$ & $20.50$ & $23.14$ & $4.02$ & $5.71$ & $6.41$
\\ 
\cline{2-14}
& Case 2 
& $1.65$ & 
$2.24$ & $2.48$
& $9.31$ & 
$12.32$ & $13.54$
& $10.80$ & $16.00$ & $18.31$ & $3.07$ & $4.53$ & $5.19$
\\ 
\cline{2-14}
& Case 3 
& $1.28$ & 
$1.80$ & $2.03$
& $7.21$ & 
$9.95$ & $11.10$
& $8.74$ & $13.14$ & $15.19$ & $2.50$ & $3.78$ & $4.36$
\\ \hline \hline
\multirow{3}[6]{0.2cm}{\rotatebox[origin=c]{90}{{\hspace{0.2in}} $\lambda=8$}}
& Case 1 
& $2.40$ & 
$2.88$ & $3.05$
& $13.44$ & 
$15.57$ & $16.32$
& $15.69$ & $20.98$ & $23.06$ & $4.25$ & $5.54$ & $6.03$
\\ 
\cline{2-14}
& Case 2 
& $1.71$ & 
$2.11$ & $2.26$
& $9.60$ & 
$11.62$ & $12.37$
& $11.79$ & $15.88$ & $17.47$ & $3.27$ & $4.37$ & $4.81$
\\ 
\cline{2-14}
& Case 3 
& $1.33$ & 
$1.69$ & $1.83$
& $7.47$ & 
$9.36$ & $10.10$
& $9.48$ & $12.81$ & $14.18$ & $2.66$ & $3.63$ & $4.02$
\\ \hline \hline
\multirow{3}[6]{0.2cm}{\rotatebox[origin=c]{90}{ {\hspace{0.2in}} $\lambda=12$}}
& Case 1 
& $2.44$ & 
$2.84$ & $2.98$
& $13.55$ & 
$15.29$ & $15.84$
& $16.90$ & $22.04$ & $24.07$ & $4.42$ & $5.55$ & $5.95$
\\ 
\cline{2-14}
& Case 2 
& $1.73$ & 
$2.07$ & $2.18$
& $9.69$ & 
$11.33$ & $11.89$
& $12.50$ & $16.24$ & $17.65$ & $3.38$ & $4.34$ & $4.69$
\\ 
\cline{2-14}
& Case 3 
& $1.35$ & 
$1.64$ & $1.74$
& $7.56$ & 
$9.07$ & $9.60$
& $9.95$ & $12.93$ & $14.07$ & $2.75$ & $3.58$ & $3.90$
\\ \hline \hline
\end{tabular}
\end{center}
\end{table*}

\begin{table*}[htb!]
\caption{Explanatory Power of the Top-$K$ Blockers}
\label{table_topK}
\begin{center}
\begin{tabular}{|l|c|  c|c|c||  c|c|c|| c|c|c|}
\hline
& & \multicolumn{3}{c||}{ $\lambda = 4$}
& \multicolumn{3}{c||} {  $\lambda = 8$ }
& \multicolumn{3}{c|}{ $\lambda = 12$}
\\ \hline
& 
Percentiles &
$50$ & $90$ & $95$ &
$50$ & $90$ & $95$ &
$50$ & $90$ & $95$
\\ \hline
\multirow{5}[6]{0.2cm}{\rotatebox[origin=c]{90}{Human}}
& Top-2 
& $64.54 \%$ & $42.16 \%$ & $38.04 \%$
& $39.12 \%$ & $27.66 \%$ & $25.31 \%$
& $29.37 \%$ & $21.45 \%$ & $19.78 \%$
\\
\cline{2-11}
& Top-3 
& $84.51 \%$ & $58.44 \%$ & $53.04 \%$
& $53.42 \%$ & $38.88 \%$ & $35.72 \%$
& $40.41 \%$ & $30.28 \%$ & $28.15 \%$
\\
\cline{2-11}
& Top-4 
& $100.00 \%$ & $72.89 \%$ & $66.16 \%$
& $65.96 \%$ & $48.86 \%$ & $45.14 \%$
& $50.20 \%$ & $38.26 \%$ & $35.75 \%$
\\
\cline{2-11}
& Top-5 
& $100.00 \%$ & $86.06 \%$ & $78.09 \%$
& $77.27 \%$ & $57.93 \%$ & $53.83 \%$
& $59.06 \%$ & $45.65 \%$ & $42.76 \%$
\\
\cline{2-11}
& Top-6 
& $100.00 \%$ & $100.00 \%$ & $89.36 \%$
& $86.94 \%$ & $66.32 \%$ & $61.85 \%$
& $67.18 \%$ & $52.50 \%$ & $49.29 \%$
\\ \hline \hline
\multirow{5}[6]{0.2cm}{\rotatebox[origin=c]{90}{Vehicular}}
& Top-2 
& $70.18 \%$ & $46.48 \%$ & $42.12 \%$
& $47.48 \%$ & $34.30 \%$ & $31.51 \%$
& $39.44 \%$ & $29.14 \%$ & $26.96 \%$
\\
\cline{2-11}
& Top-3 
& $100.00 \%$ & $64.17 \%$ & $58.18 \%$
& $63.33 \%$ & $47.59 \%$ & $44.08 \%$
& $52.94 \%$ & $40.67 \%$  & $37.86 \%$
\\
\cline{2-11}
& Top-4 
& $100.00 \%$ & $78.23 \%$ & $72.42 \%$
& $76.34 \%$ & $59.02 \%$ & $55.08 \%$
& $64.28 \%$ & $50.82 \%$ & $47.56 \%$
\\
\cline{2-11}
& Top-5 
& $100.00 \%$ & $89.72 \%$ & $85.90 \%$
& $88.29 \%$ & $69.22 \%$ & $64.86 \%$
& $73.97 \%$ & $59.78 \%$ & $56.24 \%$
\\
\cline{2-11}
& Top-6 
& $100.00 \%$ & $100.00 \%$ & $100.00 \%$
& $100.00 \%$ & $78.26 \%$ & $73.72 \%$
& $79.79 \%$ & $64.93 \%$ & $61.52 \%$
\\ \hline \hline
\end{tabular}
\end{center}
\end{table*}

\begin{figure*}[htb!]
\begin{center}
\begin{tabular}{cc}
\includegraphics[height=2.3in,width=2.8in] 
{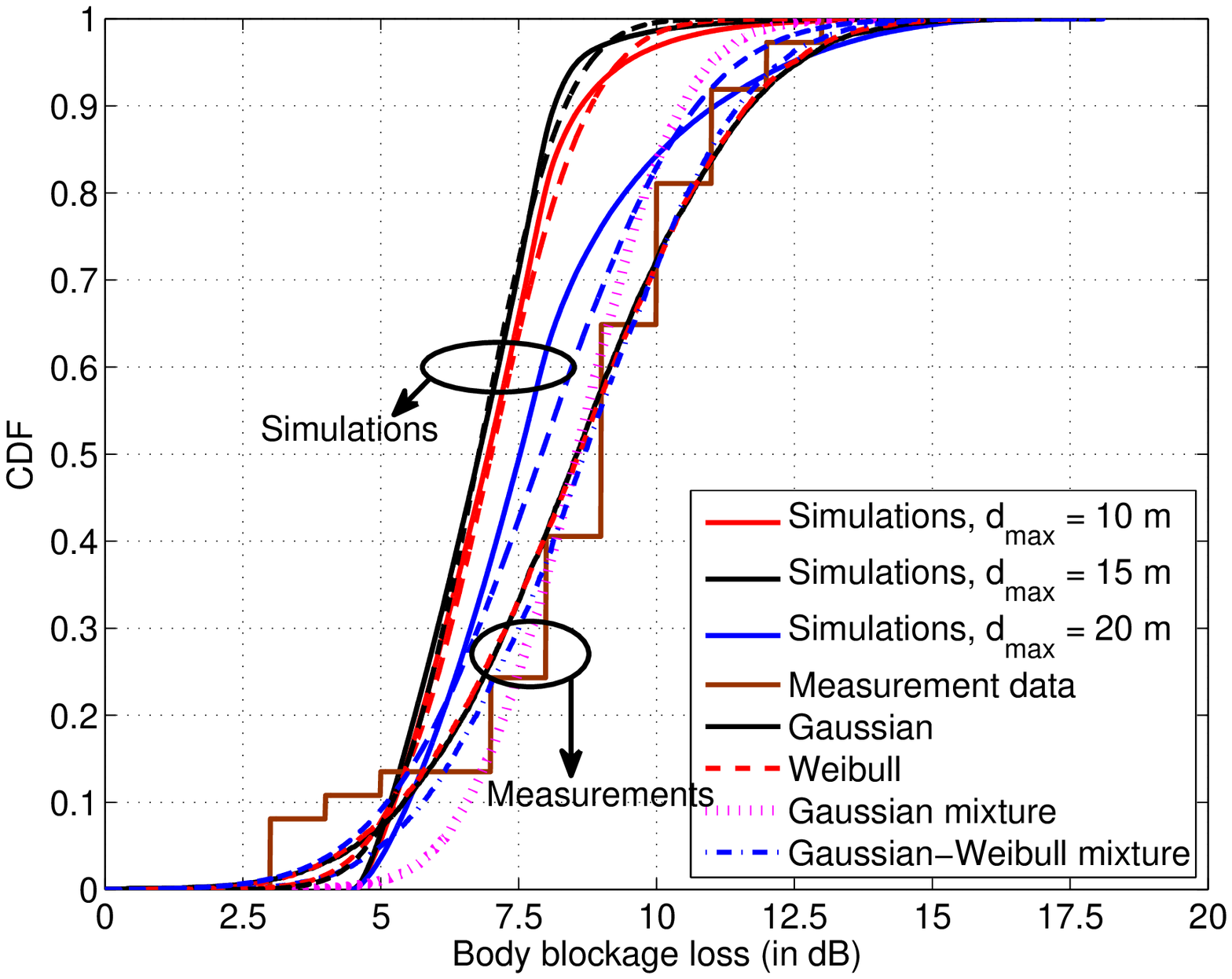}
&
\includegraphics[height=2.3in,width=3.0in]
{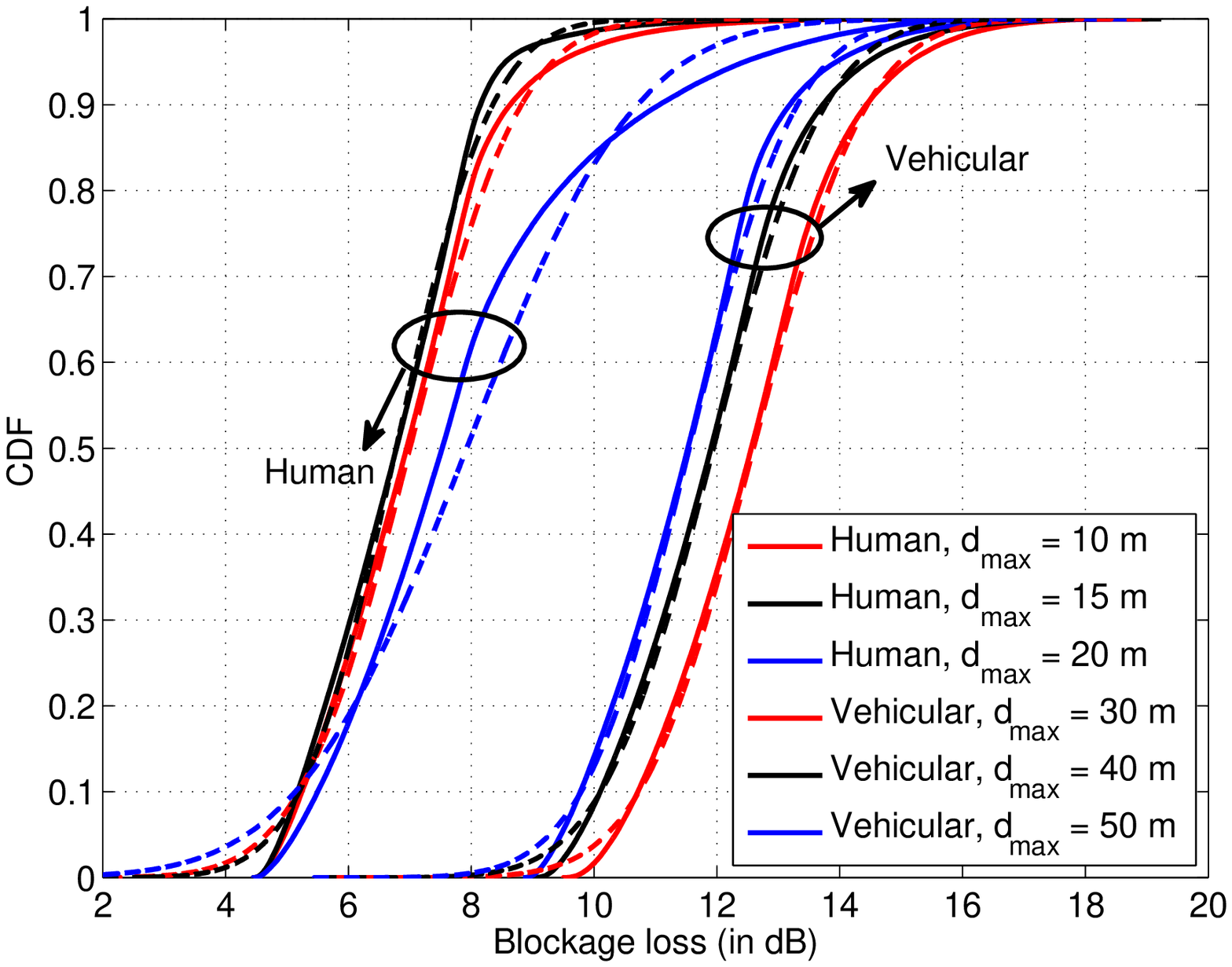}
\\ {\hspace{0.2in}}
(a) & (b)
\end{tabular}
\caption{\label{fig_scenario_humans_vehicles1}
(a) CDF of human body blockage loss with simulations using the dynamic blockage
methodology, measurements using the experimental prototype, and model fits
to measurement data. (b) CDF of human body and vehicular blockage loss with
simulations using the dynamic blockage methodology.}
\end{center}
\end{figure*}

\subsection{Loss in Signal Strength}
\label{sec3c}
For the human body blockage loss, we use the DKED model
from~\cite{5G_whitepaper,pathak,pathak2} with $d_{\sf min} = 0.5$ m, $\lambda = 4$
and the transmit-receive distance ${\sf R} = 20.5$ m. Fig.~\ref{fig_scenario_humans_vehicles1}(a)
plots the CDF of the body blockage loss (solid curves) with different
choices of $d_{\sf max}$. 
Also, plotted are Gaussian approximations to the loss estimated from the DKED
model (dashed curves). While blockage losses increase as the blockers get close to
either the transmitter or receiver, the Gaussian approximation appears to be a
reasonable first-order fit across different choices of $d_{\sf max}$.

Nevertheless, these losses are sensitive to the assumptions in the simulation
methodology described in Sec.~\ref{sec3a}. Thus, analogous to the measurements
reported in Sec.~\ref{sec2b}, link reliability studies are conducted in the third
floor of the Qualcomm building, Bridgewater, NJ with the experimental prototype
described in Sec.~\ref{sec2b} to understand the impact of (human) body blockage.
In these experiments, as illustrated in Fig.~\ref{fig_scenario_humans_vehicles}(b),
the UE side of the prototype is held stationary near the vending machine(s). The UE
antenna height is adjusted to $1$ m with possible transmission (over an NLOS link)
from one of two transmitters. One of these transmitters is at a distance of
$\approx {\hspace{0in}} 20$ m from the UE and the other is at a distance of
$\approx {\hspace{0in}} 40$ m. Both transmitters are held stationary and are
maintained at a height of $2$ m.
Except for one active subarray, beam switching across different subarrays is disabled.
Uncontrolled tests are performed where people could walk by/past the UE at
normal pedestrian speeds and at different distances (people could get as close as
$0.5$ m from the UE).

The body blockage loss\footnote{As in the self-blockage studies, RSSI could be
recorded to only within a $1$ dB precision.} corresponding to $111$ RF events with
humans blocking the UE are recorded and the CDF of this loss data is also presented
in Fig.~\ref{fig_scenario_humans_vehicles1}(a). The empirical mean and empirical
standard deviation for the loss data are $8.54$ dB and $2.45$ dB, respectively.
As with the self-blockage studies, the same set of four models (considered earlier)
are fitted to the body blockage loss data. The parameters learned in this process as
well as the KS and WKS distances between the empirical data and the fitted models
are presented in Table~\ref{table_blockage_loss_models}. From this study, we note
the best fit for the data again with a Gaussian-Weibull mixture.

From Fig.~\ref{fig_scenario_humans_vehicles1}(a), we observe that there exists a minimal yet
distinct difference between the loss estimated from simulation studies and those
with measurements. As mentioned earlier, these differences arise from the failure
of the simulation methodology to capture real deployment scenarios and addition
of more details could bridge this gap. This would be the subject of future
investigations. Nevertheless, given the lack of/difficulty in obtaining measurements
in outdoor scenarios, the simulation methodology
described in Sec.~\ref{sec3a} is considered for vehicular blockers with
$d_{\sf min} = 5$ m, $\lambda = 4$ and ${\sf R} = 100$ m. From this study,
Fig.~\ref{fig_scenario_humans_vehicles1}(b) presents a comparison between human
and vehicular blockers in terms of simulation studies. Typical median losses of
$6.5$ to $8$ dB and $11.5$ to $12.5$ dB are
seen in the human and vehicular cases, respectively. These studies illustrate
the far more significant impact vehicular blockers could have in real deployments
than human blockers making the understanding of such issues more important.

\subsection{Proposed Statistical Model}
\label{sec3d}

Table~\ref{table_proposed_model} summarizes the main conclusions of the studies
in Sec.~\ref{sec3b} and~\ref{sec3c} in terms of the number, center and angular
spread, and loss due to both human and vehicular blockers. While a
measurements-driven loss model is proposed for human body blockage, only a
simulations-based loss model (developed based on the proposed methodology in
Sec.~\ref{sec3a}) is
available for vehicular blockage. A similar model is proposed by~3GPP~to capture
blockage effects~\cite[pp.\ 53-55]{3gpp_CM_rel14_38901}. The 3GPP model differs from
the proposal in Table~\ref{table_proposed_model} in terms of the number of
blockers and their blockage regions. Further, while a {\em low-complexity}
statistical version of the DKED model is proposed for vehicular blockage
in Table~\ref{table_proposed_model}, a more explicit version is proposed for both
body and vehicular blockage in~\cite{3gpp_CM_rel14_38901}. This explicit version can
lead to a substantial complexity in 5G-NR system/link level studies. Further,
this explicit version critically relies on the DKED model, whose efficacy
in capturing loss in the human body blockage case needs further attention as
illustrated by the slight mismatch between simulation-based and measurement
studies.

\begin{table*}[htb!]
\caption{Proposed Statistical Model for Dynamic Blockage}
\label{table_proposed_model}
\begin{center}
\begin{tabular}{|c||c|c|c|c|c|}
\hline
Blocker index & $\phi_k$ & $x_k$ & $\theta_k$ & $y_k$ & Blockage loss (in dB) \\
\hline
\hline
$k = 2, 3, 4, 5$ (Human) & ${\cal U}([0^{\sf o}, \hsppp 360^{\sf o}))$
& $2.5^{\sf o}$ & $90^{\sf o}$ & $15^{\sf o}$ &
Low-complexity: 
${\cal N}(\mu = 8.5 \hspp {\sf dB}, \hspp \sigma = 2.5 \hspp {\sf dB})$
\\
& & & & & High-complexity: 
Gaussian-Weibull mixture with $p_1 = 0.15$, \\
& & & & & $\mu = 9.5$, $\sigma = 1.95$,
$p_2 = 0.85$, $\alpha = 9.4$, and $\beta = 3.7$
\\
\hline
$k = 6, 7, 8$ (Vehicular) & ${\cal U}([0^{\sf o}, \hsppp 360^{\sf o}))$
& $15^{\sf o}$ & $90^{\sf o}$ & $5^{\sf o}$ &
Simulations-based: 
${\cal N}(\mu = 12$ dB, $\hspp \sigma = 1.5$ dB)
\\
\hline \hline
\end{tabular}
\end{center}
\end{table*}

\section{Time-Scales of Blockage Events}
\label{sec4}

Understanding the time-scales at which blockage events happen can help us 
mitigate these disruptions in terms of signal quality degradation and even
possible link losses. It is important to note that these time-scales are determined
by the dynamics of blockage, and in particular, the speed at which humans walk
(or other blockers emerge and depart) to block a link or the speed at which the hand grabs the
UE and blocks the link. Towards this goal, we define the {\em link degradation time}
as the time required for the RSSI to drop from the steady-state value to its
minima in the case of a good-to-moderate channel condition, or the time required for
the RSSI to drop from the steady-state value to a complete link loss in the case of
a poor channel condition. With this definition, the link degradation time serves as the
worst-case time by which a beam switching/link adaptation procedure must be enabled to
ensure that mmW coverage remains robust, reliable and seamless.

To understand the scope of link degradation and time-scales of blockage events,
six experiments are performed with the experimental prototype described in
Sec.~\ref{sec2b}. The prototype uses a proprietary transmission frame structure where
each sub-frame is $125$ us. Analog beamforming with proprietary directional
codebooks (see design principles in~\cite{raghavan_jstsp,vasanth_comm_mag_16})
is implemented at both the base-station and UE ends. These codebooks correspond
to testing the link over $16$ transmit side beams and $20$ UE side beams ($5$
beams over four subarrays) for a beam scanning periodicity/latency of $40$ ms.
Thus, any link degradation/loss can be estimated to within an accuracy of
$\pm 20$ ms. The first four of these six experiments correspond to link
degradation due to dynamic blockage (humans walking around near the UE) and
the last two experiments correspond to self-blockage (the use of the hand).
Each experiment corresponds to different link/channel conditions with multiple
independent tests performed in these settings. More details on these experiments
including the number of tests are provided in Table~\ref{table_link_degradation}.

\begin{figure*}[htb!]
\begin{center}
\begin{tabular}{cc}
\includegraphics[height=2.3in,width=2.8in]
{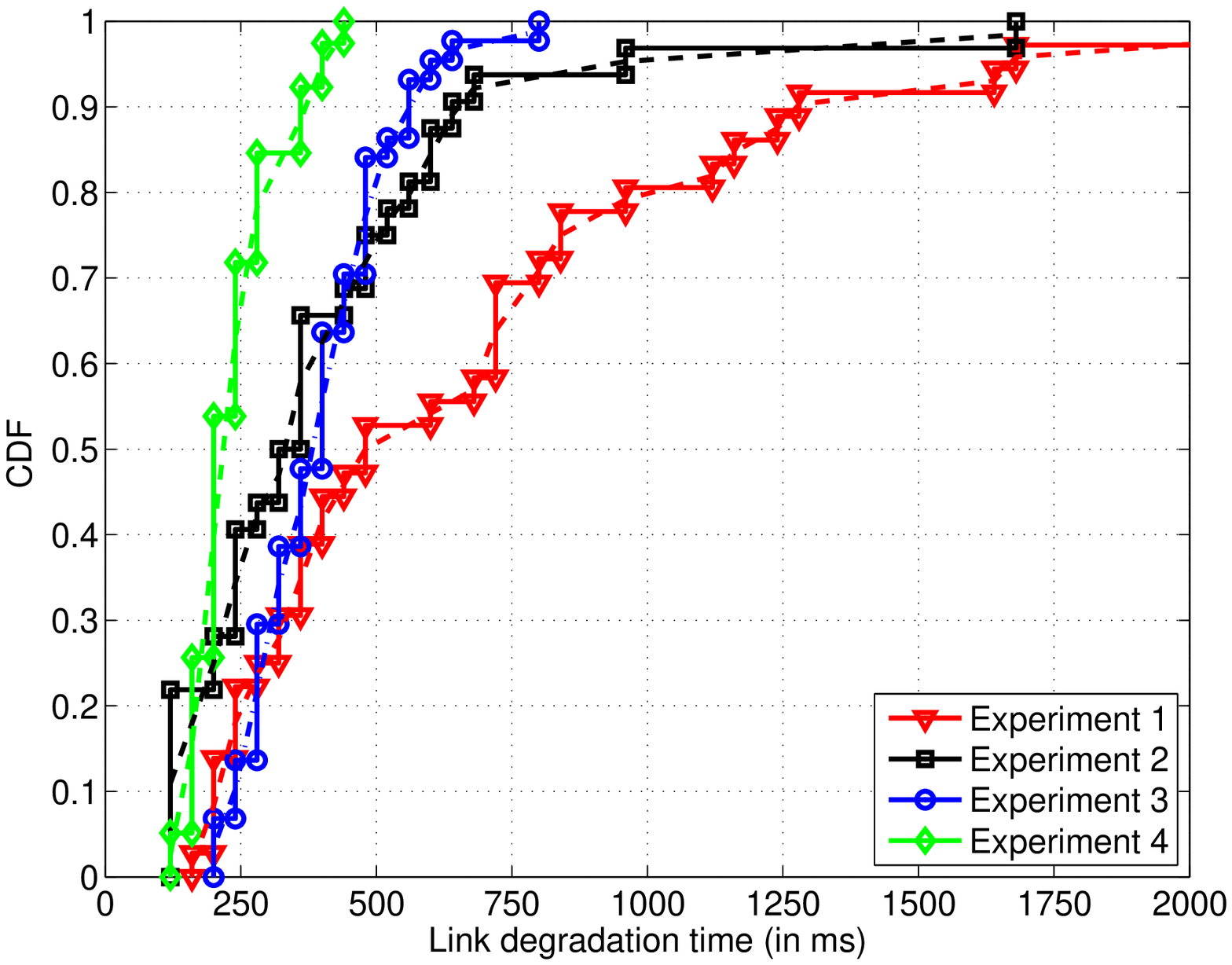}
&
\includegraphics[height=2.3in,width=3.0in]
{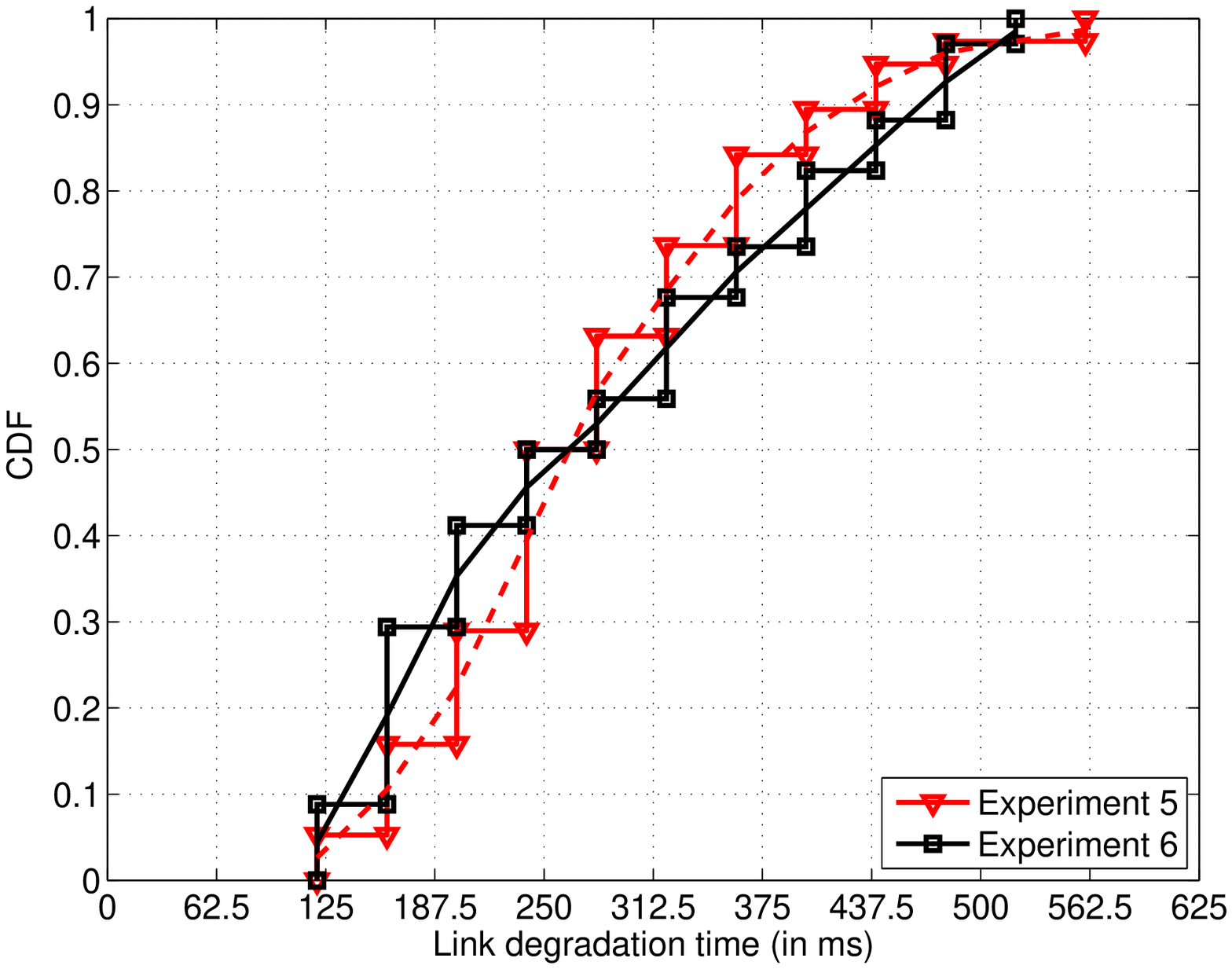}
\\ {\hspace{0.2in}}
(a) & 
(b)
\end{tabular}
\caption{\label{fig_CDFs_link_degradation}
CDF of link degradation time for (a) body blockage and (b) hand blockage
experiments.}
\end{center}
\end{figure*}

Figs.~\ref{fig_CDFs_link_degradation}(a)-(b) capture the CDF of link degradation
time across different channel conditions with human body blockage and
self-blockage, respectively. The CDFs from the true data are presented with solid lines
and piecewise linear estimated fits (across adjacent sample points) are presented
with dashed lines. From these plots, we note that the link degradation time generally
decreases as the channel condition deteriorates with no substantial difference between
hand and body blockage dynamics. Thus these plots suggest that the time-scales at which
blockages are observed at the UE end are indicative of physical
movements (of either humans or the hand) which can be on the order of a few $100$s of
ms (or slower). Thus, from Fig.~\ref{fig_CDFs_link_degradation}, it is not surprising
to see that the median value of link degradation time being on the order of
$200$-$480$ ms for body blockage and $240$ ms for hand blockage. Given the sub-ms
latencies for beam switching possible in 5G-NR, these estimates suggest that blockage
events can be handled with a robust beam management procedure.

\begin{table*}[htb!]
\caption{Description of Link Degradation Experiments}
\label{table_link_degradation}
\begin{center}
\begin{tabular}{|c||c|c|c|}
\hline
Experiment & Blockage type & Channel condition & Number of tests \\
\hline
\hline
$1$ & Body & Good & $36$ \\ \hline
$2$ & Body & Good-to-medium & $32$ \\ \hline
$3$ & Body & Medium & $44$ \\ \hline
$4$ & Body & Poor & $39$ \\ \hline
$5$ & Hand & Poor & $38$ \\ \hline
$6$ & Hand & Good & $34$ \\ \hline
\end{tabular}
\end{center}
\end{table*}

\section{Solutions to Combat Blockage}
\label{sec5}
Multiple solutions can either be individually/jointly considered to handle the
deleterious impact of performance degradation with blockage.
\begin{itemize}
\item Network densification: Beamforming design for mmW systems is expected to
leverage {\em directional} solution structures due to their robustness with
different beamforming architectures and their implementation
ease~\cite{vasanth_gcom15,raghavan_jstsp,vasanth_jsac2017}. Thus, both the base-station
and the UE are expected to steer their beams towards the dominant clusters in the
channel and blockage in these directions 
can significantly deteriorate the performance of mmW systems. In this context,
densifying the network with overlap in coverage across multiple cells~\cite{qualcomm}
can provide
higher fade margins to prevent link losses in mmW deployments. Further, the
deployment of multiple base-stations could lead to the feasibility of
different/distinct dominant paths from these base-stations to a certain UE via
distinct reflectors, scatterers or clusters thereby reducing the risk of dramatic
link degradations/failures due to blockage.

\item Subarray switching: Subarray diversity is critical at the UE end due to
the reduced spatial/angular coverage possible with an antenna at mmW frequencies
relative to sub-$6$ GHz frequencies. For example, the typical angular coverage
with a dipole/patch antenna is on the order of $90^{\sf o}$ to $120^{\sf o}$
implying the necessity of multiple subarrays as well as a careful
selection\footnote{The constraints associated with the location of camera(s),
speaker, microphone, sensors, etc.\ lead to a careful optimization of the
location of antennas in a form-factor UE design (see, e.g.,
Fig.~\ref{fig_subarray}).} of the locations of these subarrays for full
spherical coverage. Thus, coverage over the sphere is realized in a
form-factor UE design with distinct subarrays corresponding to distinct clusters
in the channel environment. This fact can be leveraged by allowing/enabling
a subarray switching procedure via beam management before an established
link degrades significantly.

In this context, mmW measurements reported in~\cite{qualcomm2} for indoor
environments (office and shopping mall) suggest that (on average) $4$-$5$
distinct clusters corresponding to distinct directions appear to be within
a power differential of $5$ dB of each other implying a reasonable level
of path diversity for indoor mmW deployments. Similarly, in outdoor mobility
tests with the prototype reported in~\cite[Sec.\ IV]{vasanth_comm_mag_16}, inter-base-station
beam switching and handover are shown to be both feasible and important with
blockages from static geographical/topographical blockages/features, foliage, etc.

\item Fall back mechanisms: With the above background, there could also be
scenarios where the network is not densified sufficiently or the channel
environment is sparse ensuring that there are no better clusters to switch to.
In these scenarios, the UE is left with little choice but to continue to use
the degraded link with some codebook enhancements (possibly proprietary from an
implementation standpoint). These enhancements could help
improve the array gain seen at the UE side by performing a maximum ratio
combining of the effective channel corresponding to the true channel and the near-field
effects of the hand. Alternately, the UE could consider fall back to legacy
carriers (such as 4G/LTE) or 5G-NR carriers such as those at sub-$6$ GHz frequencies.
\end{itemize}

\section{Concluding Remarks}
\label{sec6}
Given the prospects of accelerated deployment of 5G-NR systems, there has
been an emerging interest on a number of issues that need to be addressed
to make these systems technically viable and commercially
profitable~\cite{tap_overview1,tap_overview2}. The focus of this work is on
one such aspect: blockage of mmW signals due to the user itself (hand or
body parts) as well as humans or vehicles in the vicinity of the UE. To the
best of our knowledge, the impact of blockage as well as their implications
in terms of PHY layer mechanisms to ameliorate their impact have not been
reported for $28$ or $60$ GHz systems (two important use-cases of 5G-NR),
especially with form-factor UE designs.

In this context, our studies at $28$ and $60$ GHz show that large parts of
the spatial/angular coverage area can be blocked by the hand. This is because
the user's hand and body serve as primary obstacles in obscuring the radiation
coverage of the UE antennas with the size of the hand being large relative to
the UE. We also report measurement-driven studies of loss due to self-blockage
with a $28$ GHz experimental prototype. Our studies show that in contrast to
prior reports of $30$ to $40$ dB blockage losses, even a hard hand grip with
form-factor UEs could see significantly lower losses in the range of $5$ to
$20$ dB (with a median loss of $15$ dB). These discrepancies arise because of
the beamwidth differences between prior studies that are based on horn antenna
measurements (much smaller beamwidths) relative to form-factor phased arrays
(that could have larger beamwidths). These relative beamwidth differences
allow much higher signal energies to be radiated/captured in form-factor UEs.
In the more optimistic scenario of looser hand grips with gap between fingers,
some antenna elements can radiate/capture signal energy through the air gaps
leading to even further reduced blockage losses.

We also propose a simulation methodology for capturing the impact of dynamic
(human and vehicular) blockers in the vicinity of the UE. The spatial/angular
coverage lost is studied and the DKED model is used to estimate the loss due to
the blockers. Comparisons with measurement data of the blockage loss with human
blockers in an indoor office setting shows a reasonable first-order fit with the
data from the proposed simulation methodology. Given the difficulties in obtaining
measurement-based estimates for loss with vehicular blockers in outdoor deployments,
the proposed simulation framework could serve as a reasonable substitute for
system level studies. In this context, the proposed simulation methodology as well
as the statistical model generated from the data has already had some far-reaching
impact on channel modeling at 3GPP. In particular, Option A of the blockage model
in the 3GPP Rel.\ 14 channel modeling document~\cite[pp.\ 53-55]{3gpp_CM_rel14_38901}
is based on ideas expounded in this paper.

Another contribution of this work is in terms of understanding the time-scales at
which blockage events and their disruptions can be seen to impact the UE side. Based on
experiments with the $28$ GHz prototype, we show that these blockage events can be
attributed to physical movements and the time-scales at which these disruptions happen
are on the order of a few $100$ ms (or more). These estimates offer insights into the
feasibility of mitigation mechanisms to address blockage impairments. Given a rich
channel environment, network densification (as seen from the base-station perspective)
or the use of multiple subarrays (as seen from the UE perspective) can help given that
the 5G-NR standard allows sub-ms (or a few ms) effective latencies in beam/subarray
switching. In the case of a sparse channel environment (e.g., rural settings, highways,
etc.), mitigation mechanisms could be fall back on to legacy carriers or proprietary
codebook enhancements (on top of the steady-state beamforming codebooks used at the
UE end). The design of PHY layer enhancements to realize these ways forward could be
the subject of interesting future work in this area.

\appendix
\subsection{Parameter Learning for a Gaussian Mixture Model}
\label{app_model_GMM}
Let $\Phi_{\theta_1}(y)$ and $\Phi_{\theta_2}(y)$ denote the density functions of the two
Gaussians in the mixture corresponding to parameters $\theta_i = \{ \mu_i, \sigma_i \},
{\hspace{0.02in}} i = 1, 2$. That is,
\begin{eqnarray}
\Phi_{\theta_i}(y) = \frac{1}{ \sqrt{ 2 \pi \sigma_i^2} } \cdot
e^{ - \frac{ (y - \mu_i)^2}{2 \sigma_i^2} }, \hspp
-\infty < y < \infty
\nonumber
\end{eqnarray}
with the mixture model corresponding to a mixture probability $p_1$ given as
\begin{eqnarray}
\Phi_{\theta}(y) = \frac{p_1} {\sqrt{2 \pi \sigma_1^2} } \cdot e^{ - \frac{ (y-\mu_1)^2}
{2 \sigma_1^2} } +
\frac{ (1 - p_1)} {\sqrt{2 \pi \sigma_2^2} } \cdot
e^{ - \frac{ (y-\mu_2)^2}{2 \sigma_2^2} }.
\nonumber
\end{eqnarray}
We use ${\cal L}$ to denote the log likelihood function for the data (denoted as
$y_i, {\hspace{0.03in}} i = 1, \cdots, N$) and an unobserved latent variable,
$\Delta_i \in \{0,1 \}$. The observation $y_i$ comes from Model 1 (captured by
$\theta_1$) if $\Delta_i = 0$, or from Model 2 (captured by $\theta_2$) if
$\Delta_i = 1$. This log likelihood function is given as
\begin{eqnarray}
{\cal L} = \sum_{i = 1}^N \Big[ (1 - \Delta_i)\log \left( \Phi_{\theta_1}(y_i) \right)
+ \Delta_i \log \left( \Phi_{\theta_2}(y_i) \right) \Big]
+ \sum_{i = 1}^N \Big[ (1 - \Delta_i) \log(1 - \pi) + \Delta_i \log(\pi) \Big]
\nonumber
\end{eqnarray}
where $\pi$ denotes ${\sf P}(\Delta_i = 1)$.
Since we do not know $\Delta_i$, we replace it with its conditional expectation:
\begin{eqnarray}
\gamma_i(\theta) = {\bf E} \Big[ \Delta_i |\theta, \{y_j\} \Big] =
{\sf P}\left( \Delta_i = 1|\theta, \{y_j\} \right)
\nonumber
\end{eqnarray}
and perform the Expectation Maximization algorithm for learning the
parameters~\cite[8.5.1, pp.\ 272-275 and Algorithm 8.1]{hastie_tibshirani}. To
keep this paper self-contained, we provide the parameter learning algorithm
corresponding to stopping at $k_{\sf max}$ iterations below. The choice of
$k_{\sf max}$ is determined by an {\em a priori} choice of based on a stopping
criterion.

\begin{algorithm}
\caption{(Parameter learning for a Gaussian mixture model)}

\label{alg:parameter_learning}
\begin{algorithmic}
\State{For $k = 1$, initialize $\widehat{p}_{1,1} = 0.5$,
$\widehat{\mu}_{1,1} = \min \limits_i y_i 
$, $\widehat{\mu}_{2,1} = \max \limits_i y_i$, 
$\overline{y} = \frac{\sum_i y_i}{N}$, $\widehat{\sigma}_{1,1}^2 =
\widehat{\sigma}_{2,1}^2 = \frac{1}{N} \sum_i (y_i - \overline{y})^2$.}
\ForAll{$k = 1, \hdots, k_{\mathsf{max}}-1$}
\State\ParState{Define $\widehat{\gamma}_i = \frac{  \frac{\widehat{p}_{1,k} }
{ \sqrt{2 \pi \widehat{\sigma}_{1,k}^2}} \times
\exp \left( - \frac{( y_i - \widehat{\mu}_{1,k} )^2 }
{2  \widehat{\sigma}_{1,k}^2 } \right)
}
{   \frac{\widehat{p}_{1,k} }
{ \sqrt{2 \pi \widehat{\sigma}_{1,k}^2}}
\times \exp \left( - \frac{( y_i - \widehat{\mu}_{1,k} )^2 }
{2  \widehat{\sigma}_{1,k}^2 } \right) +
\frac{\widehat{p}_{2,k} }
{ \sqrt{2 \pi \widehat{\sigma}_{2,k}^2}}
\times \exp \left( - \frac{( y_i - \widehat{\mu}_{2,k} )^2 }
{2  \widehat{\sigma}_{2,k}^2 } \right)
}$ for $i = 1, \cdots, N$.

}

\State\ParState{Update $\widehat{\mu}_{1,k + 1}$ and $\widehat{\mu}_{2,k + 1}$
with $\frac{\sum_i \widehat{\gamma}_i y_i}{\sum_i \widehat{\gamma}_i }$ and
$\frac{\sum_i (1 - \widehat{\gamma}_i) y_i}{\sum_i (1 - \widehat{\gamma}_i) }$, respectively.
}

\State\ParState{Update $\widehat{\sigma}_{1,k + 1}^2$ and $\widehat{\sigma}_{2,k + 1}^2$
with $\frac{\sum_i \widehat{\gamma}_i ( y_i - \widehat{\mu}_{1,k+1})^2 }
{\sum_i \widehat{\gamma}_i  }$ and
$\frac{\sum_i (1 - \widehat{\gamma}_i) ( y_i - \widehat{\mu}_{2,k+1})^2 }
{\sum_i (1 - \widehat{\gamma}_i )}$, respectively.
}

\State\ParState{Update $\widehat{p}_{1,k+1}$ with $\frac{ \sum_i \widehat{\gamma}_i }{N}$.
}
\EndFor
\end{algorithmic}
\end{algorithm}

\qed

{\vspace{-0.05in}}
\bibliographystyle{IEEEbib}
\bibliography{newrefsx}

\end{document}